\documentclass[%
aps,
prx,
reprint,
superscriptaddress,
nofootinbib,
amsmath,
amssymb
]{revtex4-2}
\usepackage{
    physics,
    graphicx,
    multirow,
    tabularx,
    mathtools,
    placeins,
    nicefrac,
    hyperref,
    threeparttable,
    siunitx,
    enumitem,
    makecell
}
\usepackage{algorithm}
\usepackage{algorithmicx}
\usepackage{algpseudocode}
\usepackage[caption=false]{subfig}
\usepackage[capitalize]{cleveref}
\Crefname{section}{Sec.}{Secs.}

\usepackage[dvipsnames]{xcolor}
\hypersetup{
    colorlinks,
    linkcolor={blue!50!black},
    citecolor={blue!50!black},
    urlcolor={blue!80!black}
}

\usepackage[caption=false]{subfig}
\DeclarePairedDelimiter\pbra{\langle\!\langle}{\rvert}
\DeclarePairedDelimiter\pket{\lvert}{\rangle\!\rangle}
\DeclarePairedDelimiterX\pbraket[2]{\langle\!\langle}{\rangle\!\rangle}{#1 \delimsize\vert #2}
\DeclarePairedDelimiterX\pketbra[2]{\lvert}{\rvert}{#1 \rangle\!\rangle\!\langle\!\langle #2}

\begin{document}

\title{Noise-aware Time-optimal Quantum Control}

\newcommand{\qmaddress}{\affiliation{Quantum Motion, 9 Sterling Way, London N7 9HJ, United Kingdom}}
\newcommand{\oxddress}{\affiliation{Department of Materials, University of Oxford, Parks Road, Oxford OX1 3PH, United Kingdom}}

\author{Minjun Jeon}
\email{minjun.jeon@materials.ox.ac.uk}
\qmaddress
\oxddress

\author{Zhenyu Cai}
\email{cai.zhenyu.physics@gmail.com}
\qmaddress
\oxddress

\date{\today}

\begin{abstract}  
    Quantum optimal control plays a vital role in many quantum technologies, including quantum computation. One of the most important control parameters to optimise for is the evolution time (pulse duration). However, most existing works focus on finding the shortest evolution time theoretically possible without offering explicit pulse constructions under practical constraints like noise in the system. 
    This paper addresses these limitations by introducing an efficient method to perform the Chopped Random Basis (CRAB) optimisation in the presence of noise, specifically when the noise commutes with the gate Hamiltonian. This noise-aware approach allows for direct optimisation of the evolution time alongside other control parameters, significantly reducing the computational cost compared to full noisy simulations. The protocol is demonstrated through numerical simulations on state-to-state transfer and gate compilation problems under several noise models. Results show that the optimised fidelity has a strong dependence on evolution time due to noise, drift Hamiltonian, and local traps in optimisation, highlighting the necessity of optimising evolution time in practical settings that can lead to a substantial gain in the fidelity. Our pulse optimisation protocol can consistently reach the global optimal time and fidelity in all of our examples.  We hope that our protocol can be the start of many more works on the crucial topic of control pulse time optimisation in practical settings. 
\end{abstract}

\maketitle
\section{Introduction}
Quantum optimal control, which is the process of designing control pulses or sequences that achieve desired quantum dynamics, is central to the practical implementation of many quantum technologies such as quantum communication~\cite{PhysRevA.87.053412, Brown_2023}, quantum sensing~\cite{Titum_2021,RevModPhys.92.015004}, state preparation~\cite{li2023optimalcontrolquantumstate,Gunther_2021,Petruhanov_2022,Parajuli2023-ga,Bond_2024,Araz_2025} and gate compilations~\cite{Litteken_2023,Seifert_2023,gunther2023practicalapproachdetermineminimal,Cho_2024,gunther2023practicalapproachdetermineminimal}. Many different quantum optimal control techniques have been developed, ranging from techniques that iteratively refine control pulses based on gradients of the cost function like gradient ascent pulse engineering (GRAPE)~\cite{Khaneja_2005} and Krotov method~\cite{krotov_1983}, to techniques that try to reduce the parameter search space using chopped random basis (CRAB). Due to the advent of artificial intelligence, there are also many works trying to integrate machine learning into the pipeline~\cite{Wu_2019(DL_1), Zeng_2020(DL_2), Huang_2022(DL_3), Schafer_2020(DL_4), Ostaszewski(DL_5), Perrier(DL_6), Bukov_2018(RL_1), Zheng_2019(RL_2), Zhang_2019(RL_3), Baum_2021(RL_4), Murphy_2018(RL_5), Zheng_2021(RL_6), Andreason_2019(RL_7), Schuff_2020(RL_9)}. In addition, there were also proposals using quantum-classical methods~\cite{PhysRevResearch.5.023173, zhou2024variationalquantumcompilingthreequbit}, and such techniques have been applied to both open and close quantum systems~\cite{Petruhanov_2023(GRAPE_OPEN_1), Abdelhafez_2019(GRAPE_OPEN_2), Zeng_2020(DL_2), Ma_2015(DE_4), Wang_2024(OPEN_SYSTEM_3)}.

Pulse duration (evolution time) is one of the most important aspects in control pulse optimisation, which brings about a series of theoretical frameworks for time-optimal quantum control, for example, using Pontryagin's Maximum Principle (PMP)~\cite{Boscain_2014,Albertini_2015,Jirari2021,boscain_2021,NAGHDI2022128297, Lin_2022, PhysRevA.99.022327, Jafarizadeh2022-jk}, Quantum Brachistochrone (QB)~\cite{Carlini_2006, carlini_2007, QB, Dou_2023, morrison2023timeoptimalqubitcomputer, Jameson_2024}, geometric approach~\cite{zhang_2003, tan2024geometricoptimizationquantumcontrol, tang_2023, meinersen2024quantumgeometricprotocolsfast}, information-theoretic approach~\cite{PhysRevLett.113.010502, Muller2022-nq, Hou2023-il}, and Lie algebraic approach~\cite{Khaneja_2001, zhang_2003}. These analyses usually assume noiseless quantum systems and are interested in the theoretically achievable control pulse rather than what can be found via optimisation in practice. Under such context, they are interested in finding the minimum time required to reach a target state or perform a target gate with 100\% fidelity. Such a lower bound on the time required is also called the quantum speed limit, which has been studied in several notable numerical simulations~\cite{li2023optimalcontrolquantumstate, li2023estimationoptimalcontroltwolevel, gunther2023practicalapproachdetermineminimal} and experiments~\cite{Dong_2021, Dong2024-fr, PhysRevA.100.042315}.

However in practice, pulse optimisation can be stuck in different local traps given different pulse durations. Furthermore, there will inevitably be noise in our quantum system that can interfere with our gate Hamiltonian. Both of these factors mean that 100\% fidelity is not achievable in practice and the minimal time derived in theory to reach this perfect fidelity is not necessarily the optimal time in practice. In this article, we will try to address these limitations by devising a way to efficiently perform one of the most practical quantum optimal control protocols, CRAB, in the presence of certain noise. Such explicit inclusion of noise will allow us to perform direct optimisation for the pulse duration to maximise the fidelity reachable in practice, rather than simply searching for the theoretically possible shortest time as before. 

This article is organised as follows. In \cref{sec:TCRAB_theory}, we present a way to efficiently perform noisy CRAB under certain noise conditions and then outline our methods for performing optimisation on the pulse duration. In \cref{sec: results}, we perform numerical simulations of state-to-state transfer and gate compilation for different physical systems using our time-optimised CRAB method, which is followed by discussions on the importance of such time optimisation in \cref{sec: sensitivity}. At the end in \cref{sec:concl}, we summarise our results and list out the many interesting directions for further investigation.

\section{Time Optimisation in Quantum Optimal Control} \label{sec:TCRAB_theory}

\subsection{Chopped Random Basis (CRAB)}\label{sec:crab}

In the state-to-state transfer problem, our goal is to arrive at the target state $\ket{\psi_{g}}$ from the initial state $\ket{\psi_{0}}$, using a time-independent Hamiltonian generated by the set of basis $\{H_{i}\}^{N_{H}}_{0}$:
\begin{align}\label{eqn:Hamiltonian_TCRAB}
    H(t) = H_0 + \sum^{N_{H}}_{i=1}f_{i}(t)H_{i}.
\end{align}
This Hamiltonian is completely determined by the set of pulses, $\{f_{i}(t)\}^{N_{H}}_{i=1}$. The quantum state will evolve from the initial state following the time-dependent Schr\"{o}dinger equation with the Hamiltonian $H(t)$, giving rise to the final state, $\ket{\psi_{f}}$. In \cref{eqn:Hamiltonian_TCRAB}, the time-independent term, $H_{0}$, is called the \emph{drift Hamiltonian} while the other part is called the \emph{control Hamiltonians}.

Our goal is to find the set of pulses, $\{f_{i}(t)\}$, that maximises fidelity between the final state and the target state,  
\begin{align}
   F = \abs{\bra{\psi_{f}}\ket{\psi_{g}}}^{2}.
\end{align}

Caneva et al.\cite{Caneva_2011} developed a quantum optimal control method called the Chopped Random Basis(CRAB), where the control pulses, $\{f_{i}\}^{N_{H}}_{i=1}$, are expressed in terms of truncated Fourier basis:
\begin{align}\label{eqn:CRAB}
    f(t;\vec{\alpha},\vec{\omega}) = \alpha_0 + \sum_{m = 1}^{M} \alpha_{-m} \cos(\omega_m t) + \alpha_{m} \sin(\omega_m t),
\end{align}
where $\vec{\omega}$ is a vector of frequencies randomly drawn around the principal harmonics~\cite{Caneva_2011}. The $k$th frequency is defined as  $\omega_{k} = 2\pi k(1+r_{k})/T$, where $r_{k}$ is drawn from a uniform distribution in the range of $-0.5\leq r_{k} \leq 0.5$ and $k=1,...,M$.

Given some control pulses, $\{f_{i}\}^{N_{H}}_{i=1}$ and time duration $T$, the control unitary becomes
\begin{align}\label{eqn:control_unitary_continuous}
    U(T,\vec{\alpha}) = \mathcal{T} \exp{-i\int_{0}^{T} dt (H_0 + \sum^{N_{H}}_{i=1}f_{i}(t;\vec{\alpha}_i,\vec{\omega}_i) H_{i})}.
\end{align}
In some of the experiments in Ref.~\cite{Caneva_2011}, the evolution time  T  was chosen to be inversely proportional to the energy scale, with an arbitrarily selected constant. In some other experiments in Ref.~\cite{Caneva_2011}, $T$ is fixed to be twice the minimum time set by the quantum speed limit. This leaves only $\vec{\alpha}$ as free parameters for optimisation, i.e. $U(T, \Vec{\alpha}) \rightarrow U(\Vec{\alpha})$.

In the absence of noise, the final output state is $\ket{\psi_{f}} =U(\vec{\alpha})\ket{\psi_{0}}$ and its state fidelity with respect to the target state is
\begin{align}\label{eqn:final_fidelity}
    F_{U}(\vec{\alpha}) = \abs{\bra{\psi_{g}}U(\vec{\alpha})\ket{\psi_{0}}}^2 = \abs{\bra{\psi_{g}}\ket{\psi_{f}}}^{2}.
\end{align}
CRAB uses the fidelity $1-F_{U}(\vec{\alpha})$ as a cost function to optimise the free parameters, $\vec{\alpha}$, often with additional constraints on the parameter depending on the problem.

\subsection{Noisy Simulation of CRAB}\label{sec:noisy_crab}
So far we have not considered noise in the quantum system, but noise is unavoidable in practice.
Directly trying to incorporate noise into the CRAB optimisation will simply make the simulation exponentially more expensive with respect to the number of qubits, since a $N$-qubit noisy mixed state simulation is equivalent to an $2N$-qubit pure state simulation. In the presence of Markovian noise, instead of following the time-dependent Schr\"{o}dinger equation, the evolution of the state will follow the Lindblad master equation
\begin{align}\label{eqn:Lindblad_master_equation}
    \dv{t}\rho &= \underbrace{- i[H, \rho]}_{\text{unitary part}} + \underbrace{\sum_{k = 1}^{4^{N}-1} \gamma_k \left(L_{k} \rho L_{k}^\dagger - \frac{1}{2}\left\{L_{k}^\dagger L_{k}, \rho\right\}\right)}_{\text{dissipative part}}
\end{align}
where $\{L_k\}$ are the jump operators that describe the noise process. As explicitly shown in \cref{sec:commute_noise}, we can vectorise the density operator to write the Lindblad master equation in the Liouville superoperator form~\cite{manzanoShortIntroductionLindblad2020}:
\begin{align}\label{eqn:Liouville}
    \dv{t} \pket{\rho} = \mathcal{L} \pket{\rho} = \left(\mathcal{L}_H + \mathcal{L}_D\right) \pket{\rho}
\end{align}
where $\mathcal{L}_H$ represent the Liouville operator of the unitary part and $\mathcal{L}_D$ represent the Liouville operator of the dissipative part. For simplicity, we will consider the case in which both $\mathcal{L}_H$ and $\mathcal{L}_D$ are time-independent. In this case, with an incoming state $\rho_0 = \ketbra{\psi_0}$, the resultant noisy state at time $T$ is simply given by:
\begin{align*}
    \pket{\rho_{f,\mathrm{noi}}} = e^{\left(\mathcal{L}_H + \mathcal{L}_D\right)T}\pket{\rho_0}
\end{align*}
and its fidelity against the target \emph{pure} state $\pket{\rho_g} = \ketbra{\psi_{g}}$ is given as
\begin{align*}
    \Tr(\rho_g\rho_{f,\mathrm{noi}}) = \pbraket{\rho_g}{\rho_{f,\mathrm{noi}}} = \pbra{\rho_g}e^{\left(\mathcal{L}_H + \mathcal{L}_D\right)T}\pket{\rho_0}.
\end{align*}
Evaluating this fidelity requires full simulation of mixed state vectors of dimension $4^{N}$ over many time steps, which as mentioned, is exponentially more expensive than the pure state simulation of dimension $2^{N}$ required for the noiseless case in \cref{eqn:final_fidelity}.

In order to reduce the computational cost, we will consider the case in which the unitary part and the dissipative part commute. As shown in \cref{sec:commute_noise}, a sufficient condition is
\begin{align}\label{eqn:commutation_condition}
    \left[H, L_k\right] = a_k L_k \quad \forall k \quad \Rightarrow \quad \left[\mathcal{L}_H, \mathcal{L}_D\right]
\end{align}
for some set of real number $\{a_k\}$. Physically, this means that the jump operator $L_k$ will map one eigenvector of $H$ to another. When the jump operators are Pauli operators, they will generate Pauli noise channels that are diagonal in the Pauli transfer matrix formalism~\cite{greenbaumIntroductionQuantumGate2015}. For such Pauli noise, another (not mutually exclusive) way for the unitary and dissipative part to commute is to have $\mathcal{L}_H$ block diagonal in the same way as the degenerate subspaces of $\mathcal{L}_D$ as discussed in \cref{sec:pauli_channel_Lindblad}.

When $\mathcal{L}_H$ and $\mathcal{L}_D$ commutes, the output fidelity can be written as
\begin{align}
    \pbraket{\rho_g}{\rho_{f,\mathrm{noi}}} = \pbra{\rho_g}e^{\mathcal{L}_D T}e^{\mathcal{L}_H T}\pket{\rho_0} = \pbraket{\rho_{g,\mathrm{noi}}}{\rho_{f}}
\end{align}
with
\begin{align}
    \pket{\rho_f} &= e^{\mathcal{L}_H T}\pket{\rho_0}\\
    \pket{\rho_{g,\mathrm{noi}}} &= \left(e^{\mathcal{L}_D T}\right)^\dagger \pket{\rho_{g}}.
\end{align}
Here $\rho_f = \ketbra{\psi_f}$ is the noiseless final state we have before. Note that we have assumed that $\mathcal{L}_H$ is time-independent so far, but the same expression $\pbraket{\rho_g}{\rho_{f,\mathrm{noi}}} = \pbraket{\rho_{g,\mathrm{noi}}}{\rho_{f}}$ is obtained even if $\mathcal{L}_H$ is time-dependent, with the only change that $\pket{\rho_f}$ is now a state dependent on the pulse parameters $\vec{\alpha}$ as described in \cref{sec:crab}. The condition in \cref{eqn:commutation_condition} needs to hold for all $t$, but in practice, we simply check in against all of the subterms in the Hamiltonian in \cref{eqn:Hamiltonian_TCRAB}. If the jump operators are Hermitian or anti-Hermitian, then $e^{\mathcal{L}_D T}$ will be self-adjoint and thus we have $\pket{\rho_{g,\mathrm{noi}}} = e^{\mathcal{L}_D T} \pket{\rho_{g}}$ being simply the noisy target state that undergoes the same noise channel.

Hence, we can obtain an estimate of the noisy fidelity by simply performing $2^N$-dimensional pure state simulation in the same way as in \cref{sec:crab} to obtain the noiseless output state $\rho_{f} = \ketbra{\psi_f}$, then we can obtain the noisy fidelity by measuring the modified observable $\rho_{g,\mathrm{noi}}$ on the noiseless state. The form of the observable $\rho_{g,\mathrm{noi}}$ is independent of the control pulses and thus can be calculated beforehand before all of the pulse optimisations. As shown in \cref{sec:pauli_channel}, under Pauli noise, we can write out the exact $T$-dependence for observable $\pbra{\rho_{g,\mathrm{noi}}(T)}$
\begin{align*}
    \pbra{\rho_{g,\mathrm{noi}}(T)} = 2^{-N}\sum_{j} e^{- \lambda_{j} T} \pbraket{\rho_{g}}{G_j} \pbra{G_j},
\end{align*}
which allows for a simpler calculation of $\rho_{g,\mathrm{noi}}(T)$ at different $T$.
Here $\{G_j\}$ is the Pauli basis, and $\lambda_j$ is a real number determined by the set of jump operators that anti-commute with $G_j$. We can further simplify the sum above by truncating it to include only terms with significant value of $e^{- \lambda_{j} T} \pbraket{\rho_{g}}{G_j}$. Performing simulation in the way outlined above is significantly cheaper than performing $4^{N}$-dimension noisy simulation using the Lindblad master equation through all the time steps to obtain $\rho_{f,\mathrm{noi}}$ for every iteration of pulse optimisation.

Using the expression of $\pbra{\rho_{g,\mathrm{noi}}(T)} $ for Pauli noise above, we see that the output fidelity will decay in a multi-exponential manner
\begin{align*}
    & F (T, \vec{\alpha}) = \pbraket{\rho_g}{\rho_{f,\mathrm{noi}} (T, \vec{\alpha})} = \pbraket{\rho_{g,\mathrm{noi}}(T)}{\rho_{f}  (T, \vec{\alpha})} \\
    &= 2^{-N}\sum_{j} e^{- \lambda_{j} T} \pbraket{\rho_{g}}{G_j} \pbraket{G_j}{\rho_{f} (T, \vec{\alpha})} 
\end{align*}
where we have written out explicitly the $T$ and $\vec{\alpha}$ dependence of the different components. In practice, many of these $\lambda_j$ can share very similar values, enabling us to group many of these decay terms. In particular, we have shown in \cref{sec:pauli_channel} that for a particular type of Pauli channel we call group channel~\cite{caiMultiexponentialErrorExtrapolation2021}, the fidelity will decay with a single exponential curve. One example of such a group channel is the global depolarising channel, whose fidelity decay follows 
\begin{align}
\label{eqn:fidelity_depolarising}
    F (T, \vec{\alpha}) &= \pbraket{\rho_g}{\rho_{f,\mathrm{noi}} (T, \vec{\alpha})} \nonumber \\
    &= e^{-\lambda T} F_U(T, \vec{\alpha}) + 2^{-N}(1-e^{-\lambda T}).    
\end{align}
with $F_U(T, \vec{\alpha})$ being the noiseless fidelity given in \cref{eqn:final_fidelity}.

\subsection{Implementation of Time-optimised CRAB}
After being able to more efficiently implement CRAB in the presence of noise, the natural competition between the noise, which favours shorter evolution time, and the quantum speed limit, which favours longer evolution time, will call for the need to optimise along the time direction. This brings us to time-optimised CRAB (TCRAB) in which we try to maximise $F (T, \vec{\alpha})$ over both $T$ and $\vec{\alpha}$. The first possibility is to optimise $T$ and $\vec{\alpha}$ in separate and alternating rounds. However, as shown in \cref{sec:implementation_details}, the $T$ optimisation performed after full $\vec{\alpha}$ optimisation tends to get stuck in local minima. Hence, we instead turn to a global optimiser called basin-hopping, for simultaneous optimisation of $T$ and all parameters in $\Vec{\alpha}$. Basin-hopping is a two-step optimisation method combining global search and local optimisation, ideal for rugged, funnel-shaped energy landscapes~\cite{Olson_2012(BasinHopping)}. L-BFGS-B~\cite{Zhu_1997(L-BFGS-B)}, a variant of limited-memory BFGS~\cite{Liu_1989}, was used as the local optimiser in our case.

It is also possible to perform TCRAB using root-finding methods. With a fixed evolution time $T$, we can apply CRAB to obtain the optimised parameters $\vec{\alpha}_T$ that achieve the highest possible fidelity for the given $T$
\begin{align} \label{eqn:crab_infid}
    F_{\mathrm{opt}}(T) = F(T, \vec{\alpha}_T)
\end{align}
Hence, finding the optimal evolution time is simply identifying the maxima in $F_{\mathrm{opt}}(T)$, which can also be solved by performing root-finding methods on its derivative $\dot{F}_{\mathrm{opt}}(T)$. The derivative here can be estimated using finite difference. In this article, the \emph{bisection method} is used as an example of root-finding methods to find the optimal evolution time. Using root-finding methods will return a maximum of $F_{\mathrm{opt}}(T)$, but it is not necessarily the global maximum. However, as we will see in our examples later, some $F_{\mathrm{opt}}(T)$ are actually concave, allowing us to obtain the global maximum using the bisection method, while in many other cases, we can reach a local minimum that still has very high fidelity $F_{\mathrm{opt}}$ close to the global maximum. The detailed implementations and hyper-parameters used in our simulations are all outlined in \cref{sec:numerical_simulations}.

\section{Numerical Simulations} \label{sec: results}

\subsection{State-to-State Transfer} \label{subsec: results:state-to-state-transfer}

\subsubsection{Entanglement Generation} \label{subsubsec: results:state-to-state-transfer: JosephsonChargeQubits}
We will present our simulation results to benchmark CRAB and TCRAB for a series of state-to-state transfer and gate compilation tasks. The first example is an entanglement generation for the two capacitively coupled Josephson charge qubits in a depolarising channel. The two qubits are initialised as $\ket{00}$, and the target state is set to be a bell pair, $\ket{\Psi^{+}}=(\ket{00} + \ket{11}$)/2.

As noted in Caneva et al.\cite{Caneva_2011}, the Hamiltonian of two capacitively coupled Josephson charge qubits is 

\begin{align}\label{eqn:Josephson_Hamiltonian}
    H(t) = \sum_{i=1,2}(E_{C}\sigma^{z}_{i} + E_{J}\sigma^{x}_{i}) + E_{cc}(t)\sigma^{z}_{1}\sigma^{z}_{2}. 
\end{align}
We set $E_{J} = -E_{C} = 1$ such that the energy is expressed in the units of $E_{J}$. The control Hamiltonian is $\sigma^{z}_{1}\sigma^{z}_{2}$, and the corresponding control pulse is $E_{cc}(t)$, which is expressed as a truncated Fourier series (\cref{eqn:CRAB}) parametrised by the set of parameters $\vec{\alpha}$ in CRAB and TCRAB.

We perform CRAB and TCRAB for the state-to-state transfer problem with the hyper-parameters specified in \cref{subsec: hyper-parameters} with $8$ frequencies for the basis functions, i.e. $M=8$. We will assume depolarising noise here with a decay rate $\lambda = 0.01$, which means the fidelity calculation will follow \cref{eqn:fidelity_depolarising}. 

\begin{figure*}[htbp]
    \centering
    \subfloat[\label{fig:cost_func_bell_state}]{\includegraphics[width=0.47\textwidth]{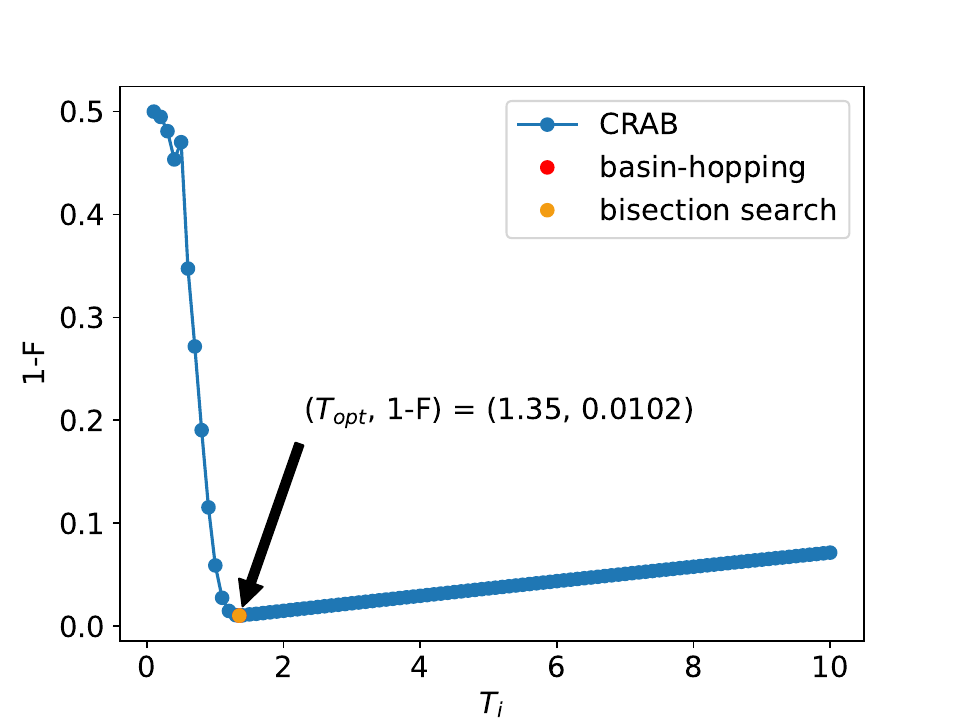}}
    \subfloat[\label{fig:cost_func_lmg}]{\includegraphics[width=0.47\textwidth]{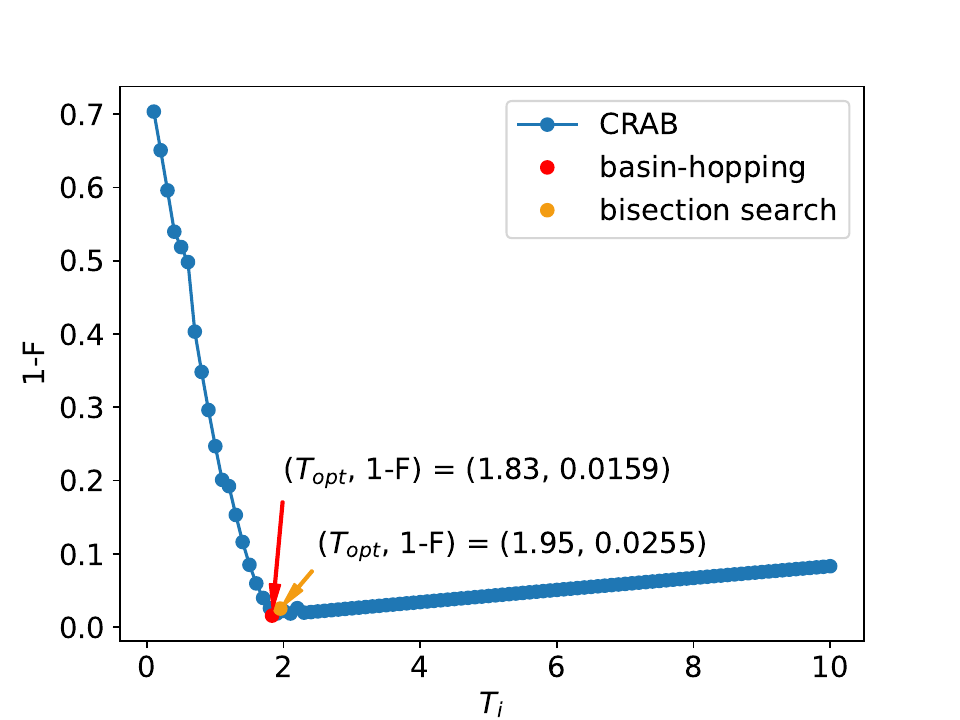}}\\
    \subfloat[\label{fig:cost_func_swap_1.0_M_8}]{\includegraphics[width=0.47\textwidth]{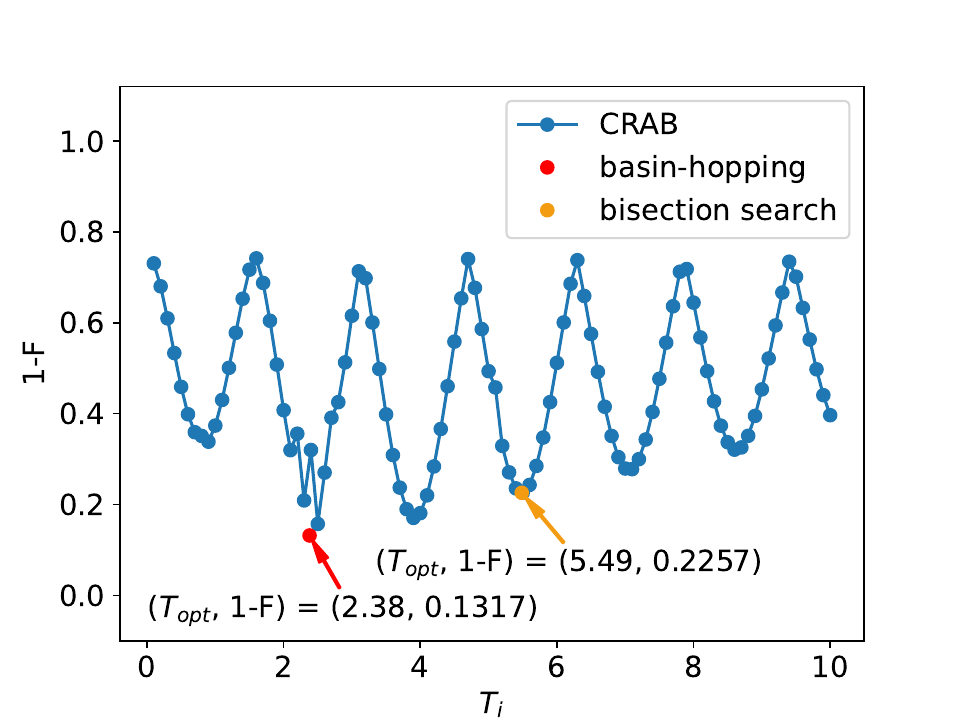}}
    \subfloat[\label{fig:cost_func_dipole_0.5_M_8}]{\includegraphics[width=0.47\textwidth]{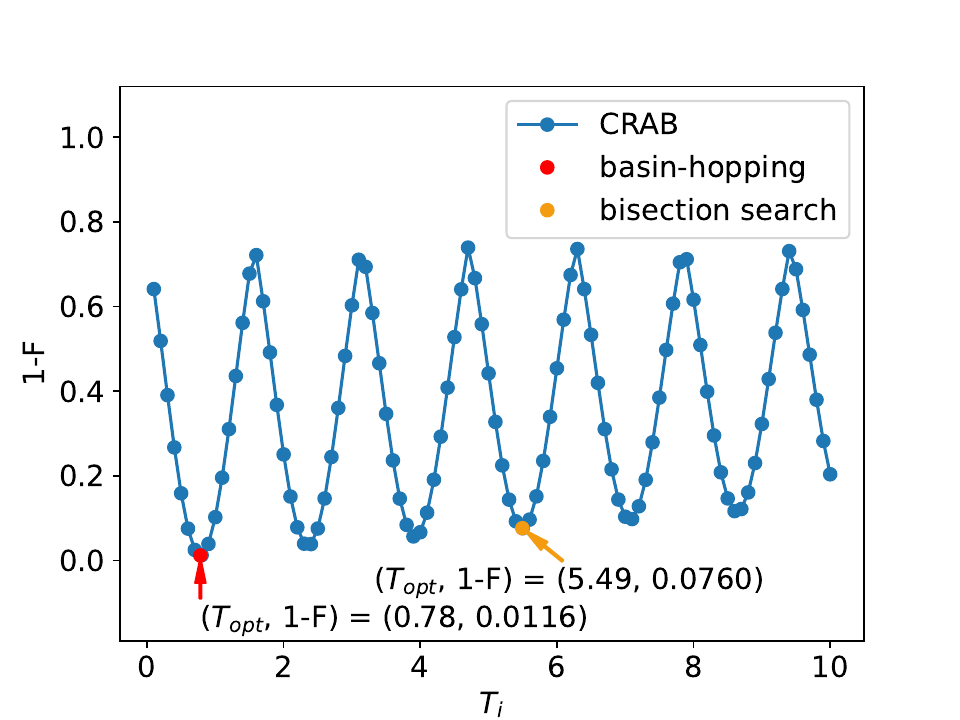}}

    \caption{The optimised infidelity reached using CRAB (blue) and TCRAB with two optimisation methods, i.e. basin-hopping (red) and bisection search (orange) for (a) entanglement generation of two capacitively coupled Josephson charge qubits; (b) state-to-state transfer from the ground state of paramagnetic phase to a ground state of ferromagnetic phase; (c) CZ gate compilation for spin qubits with SWAP control fields; (d) CZ gate compilation for spin qubits with dipole-dipole control fields. The optimal time, $T_{opt}$, and optimised infidelity are annotated. In (a), we have basin-hopping and the bisection method both successfully converged to the same global minimum since the cost function is convex. In (b), (c) and (d), due to the presence of oscillations, the bisection search converged to a local minimum instead. Note that there is a small oscillation in the cost function near $T_{i}=2$ for (b).} 
\label{fig:cost_func}
\end{figure*}

In \cref{fig:cost_func_bell_state}, we have plotted $1-F_{\mathrm{opt}}(T)$ (see \cref{eqn:crab_infid}), which is the optimal infidelity achieved by CRAB for different evolution time. Indeed, as expected, $1-F_{\mathrm{opt}}(T)$ decreases rapidly at the beginning due to the quantum speed limit, reaches an optimal point and then rises again due to noise in the evolution. 
Because of the discretisation of the time step, we are not able to read off the exact optimal time from this curve. 
We then perform TCRAB using the basin-hopping algorithm using $100$ different initial guesses of the evolution time, evenly distributed across the whole time range. The lowest infidelity achieved is $0.0102$ at the evolution time $T_{opt}=1.35$. In $72$ out of the $100$ runs, our algorithm can converge around this optimal point, outputting $T_{opt} \in [1.349, 1.359]$. We have only shown the optimal points in the plot, but more results for the rest of the runs can be found in \cref{sec:add_numerics}. 

We also perform the bisection method to search for the optimal time, with the gradient of $F_{\mathrm{opt}}$ estimated using finite difference. We are able to also obtain the same optimal evolution time $T_{opt}=1.35$ using $48$ evaluations of $F_{\mathrm{opt}}$ at different $T$ during the algorithm.

\subsubsection{Lipkin-Meshkov-Glick Model} \label{subsubsec: results:state-to-state-transfer: LMG}
The second example that we will look at is the Lipkin-Meshkov-Glick (LMG) model, which describes the uniform spin-spin interaction in the presence of a transverse magnetic field in z-direction:
\begin{align}\label{eqn:LMG_model}
    H = -\frac{J}{N} \sum_{i < j} {\sigma^{x}_{i}\sigma^{x}_{j} + \gamma \sigma^{y}_{i}\sigma^{y}_{j}} -\Gamma(t) \sum^{N}_{i=1} \sigma^{z}_{i}.
\end{align}
Here $J$ is the coupling strength of the spin-spin interactions, $N$ is the number of spins, and $\gamma$ governs the anisotropy of the spin-spin interaction. We are assuming that we can control the strength of the magnetic field, i.e. $\Gamma(t)$ is our control pulse. In the thermodynamic limit, $N \rightarrow \infty$, a second-order phase transition occurs at $\Gamma_{c} = 1$ from a ferromagnet($\Gamma < 1$) to a paramagnet($\Gamma > 1$). Here, looking at $N =3$, we will perform a state-to-state transfer from the ground state of the paramagnet ($\Gamma \gg 1$) to the ground state of the ferromagnet ($\Gamma = 0$). While there is only one ground state at the paramagnetic phase, i.e. all spins pointing in the $-z$-direction, the system has degenerate ground states at the ferromagnetic phase. Given the form of the Hamiltonian and control fields, we chose $\frac{1}{2} (\ket{000} + \ket{011} + \ket{101} + \ket{110})$ as the target state.

We perform the benchmark of CRAB and TCRAB with the hyper-parameters specified in \cref{subsec: hyper-parameters} with $10$ frequencies for the control pulse basis functions, i.e. $M=10$. We will again assume the noise here is depolarising noise with a decay rate $\lambda = 0.01$, which means the fidelity calculation follows \cref{eqn:fidelity_depolarising}.

The results are shown in \cref{fig:cost_func_lmg}. Similar to the last example of entanglement generation, the optimised infidelity of CRAB decreases sharply until the minimum point, and then increases again due to decoherence. Similar to before, we perform TCRAB with basin-hopping using $100$ different initial guesses of the evolution time. The lowest infidelity achieved is $1 - F_{opt} = 0.0160$ with the corresponding evolution time being $T_{opt}=1.83$. In $72$ out of the $100$ runs, our algorithm can converge around this optimal point, outputting $T_{opt} \in [1.819, 1.859]$ (see \cref{sec:add_numerics}).

Using the bisection method instead, we obtain the optimal evolution time $T_{opt}=1.95$ with the infidelity $0.0255$ using $34$ evaluations of $F_{\mathrm{opt}}$. We are not able to reach the exact minimum in this case due to the small oscillation of $ 1 - F_{opt}(T)$ around the optimal evolution time as can be seen in \cref{fig:cost_func_lmg}.

\subsection{Gate Compilation} \label{subsec: results:gate_compilation}
In this section, we will perform gate compilation for CZ gates between two spin qubits in quantum dots. In the lab frame, the general expression of Hamiltonian for two spin-1/2 particles in a uniform magnetic field is:
\begin{align} \label{eqn:general_spin_spin_Hamiltonian}
    H = \frac{1}{2} (E_{1}Z_{1} + E_{2}Z_{2}) +\frac{J}{2}\text{SWAP},
\end{align}
where $E_{1}$ and $E_{2}$ are Zeeman splitting of the two spin qubits, respectively. It can be rearranged into
\begin{align} \label{eqn:spin_spin_Hamiltonian_rearranged}
    H = \frac{E_{Z}}{2}(Z_{1} + Z_{2}) +\frac{\Omega}{2}(Z_{1} - Z_{2}) + \frac{J}{2}\text{SWAP},
\end{align} 
where $E_{Z}$ is the average Zeeman splitting $E_{Z} = (E_{1} + E_{2})/2$, and $\Omega$ is half of the difference between the Zeeman splitting of the two dots, i.e. $\Omega = (E_{1} - E_{2})/2$. Since the exchange interaction between two qubits can be controlled electrically by changing the plunger gate voltage, $J(t)$ will be tuneable and the related terms become our control Hamiltonian.

The main noise source in spin qubits in quantum dots is the charge noise in the various control lines~\cite{burkardSemiconductorSpinQubits2023}, which can lead to fluctuation in $J(t)$ and/or $E_{1/2}$. These will be the sources of noise that we will consider later. 

To perform the gate-compilation optimisation, we will map it into a state-to-state transfer problem using the Choi-Jamio\'{l}kowski isomorphism as further outlined in \cref{subsubsec:choi_state}. In this way, we could utilise tools we developed for the state-to-state problem in \cref{sec:TCRAB_theory} to perform gate compilations in the presence of noise.

Depending on the natural set-up of the quantum dots, which can bring about different $\Omega$, we will be interested in two different parameter regimes: $\Omega \ll J$ and $\Omega \gg J$ as will be discussed in the following sections.

\subsubsection{CZ compilation at $\Omega \ll J$ }\label{subsubsec:DD_noise}
In the regime of $\Omega \ll J$, i.e. the Zeeman splitting gradient is much smaller than the exchange interaction, the effective Hamiltonian in the rotating frame of reference is reduced to~\cite{Cai_2019}:
\begin{align}
    H &= \frac{1}{2} (\Delta E_{1}Z_{1} + \Delta E_{2}Z_{2}) +\frac{J(t)}{2} \text{SWAP} \label{eqn:rotating_frame_spin_spin_Hamiltonian}.
\end{align}
Here $\Delta E_{1}$ and $\Delta E_{2}$ are additional Zeeman splitting on top of $E_{1}$ and $E_{2}$, for example, due to micromagnets or local Stark shifts. We will assume these additional splittings to be fixed in our gate compilation. Hence, the drift Hamiltonian will lead to local $Z$ rotations, and the control Hamiltonian is the SWAP operation.

Fluctuation in the gate voltages on the quantum dot can lead to fluctuation of $\Delta E_{1}$ and $\Delta E_{2}$, which effectively becomes local dephasing channels on each qubit. Such noise channel commutes with the Hamiltonian in \cref{eqn:rotating_frame_spin_spin_Hamiltonian}, thus the final fidelity between the final and the target state can be derived using the simulation method in \cref{sec:noisy_crab} (See \cref{sec:dephasing_noise} for more details).

We will perform the gate compilation of the CZ gate using CRAB and TCRAB. The Zeeman splittings in the drift Hamiltonian were set asymmetrically: $\Delta E_{1}=1.5$, $\Delta E_{2}=0.5$. The number of basis functions was set to $M=8$, and other hyper-parameters were chosen as stated in \cref{subsec: hyper-parameters}. The local dephasing rate is set to be $0.05$ (this is the strength of the related jump operator with its definition given in \cref{sec:dephasing_noise}).

The result is shown in \cref{fig:cost_func_swap_1.0_M_8}, we see that a key difference from our previous examples is the oscillation in the infidelity. This is because the drift Hamiltonian and control Hamiltonian commute in this case. Thus, the evolution operator of the drift field, i.e. $e^{-iH_{0}T}$ leads to rotation on the multi-qubit Bloch sphere, causing oscillation in the fidelity. We will discuss such oscillations in further detail in \cref{sec: sensitivity}.

On top of oscillation, the effects due to the quantum speed limit and decoherence from the noise channel lead to an envelope resembling what we have before. In order to determine the exact optimal time, we perform TCRAB using basin-hopping and found $T_{opt} = 2.38$ to be the optimal time of evolution, which results in the lowest infidelity of $0.1317$. As further detailed in \cref{sec:add_numerics}, for the $100$ rounds of basin-hopping optimisation we perform, $13$ of them end in the right basin and give us the optimal time. Within these $13$ runs, $10$ of them start far away from the optimal time, showing that our algorithm is not susceptible to local traps. For the rest of the runs, the majority of them end in the second and third most optimal time.

Using the bisection method instead, we can obtain the optimal evolution time $T_{opt}=5.49$ with the infidelity $0.2257$ using $34$ evaluations of $F_{\mathrm{opt}}$. We have reached the third lowest basin with still very low infidelity.

\subsubsection{CZ compilation at $\Omega \gg J$}\label{subsubsec:global_z_noise}
In the regime of $\Omega \gg J$, i.e. the Zeeman splitting gradient is much larger than the exchange interaction, the effective Hamiltonian in the rotating frame of reference is reduced to:
\begin{align}
    H &= \frac{1}{2} (\Delta E_{1}Z_{1} + \Delta E_{2}Z_{2}) + \frac{J(t)}{2}Z_{1} \otimes Z_{2}\label{eqn:rotating_frame_spin_spin_Hamiltonian_first_order_approx}.
\end{align}
In this regime, the control Hamiltonian becomes the dipole-dipole interaction, i.e. the $Z \otimes Z$ term, while the drift Hamiltonian is the sum of two single Z gates, i.e. $H_{0} = \frac{1}{2} (\Delta E_{1}Z_{1} + \Delta E_{2}Z_{2})$, as in \cref{eqn:rotating_frame_spin_spin_Hamiltonian_first_order_approx}.

Here let us investigate another possible noise source coming from the oscillation of $J(t)$, which will lead to the dipole-dipole noise channels (See  \cref{appendix:subsec:dipole_dipole_channel} for more details.). Again such noise channel commutes with the Hamiltonian in \cref{eqn:rotating_frame_spin_spin_Hamiltonian_first_order_approx}, thus the final fidelity between the final and the target state can be derived using the simulation method in \cref{sec:noisy_crab}.

\cref{fig:dipole_0.5_M_8} shows the results of the gate compilation of the CZ gate using CRAB and TCRAB. We chose to use $8$ frequencies, i.e. $M=8$, to define the control pulse. The Zeeman splittings for the two quantum dots were set symmetrically: $\Delta E_{1}=1.0$, $\Delta E_{2}=1.0$. The decay rate of the dipole-dipole noise channel was set as $0.03$ (See \cref{appendix:subsec:dipole_dipole_channel}). 

Again, we see oscillation in the optimised infidelity for the same reason and there is again an envelope due to fidelity decay caused by noise. After 100 runs of TCRAB using basin-hopping, we identified $T_{opt}=0.78$ to be the optimal time of evolution, which resulted in the lowest infidelity of $0.0116$. Out of the $100$ TCRAB runs, $86$ converges to around the global optimal time. For the rest of the runs, $13$ of them converge to the second lowest basin and one run converges to the third lowest basin. Using the bisection method instead, we can obtain the optimal evolution time $T_{opt}=5.49$ with the infidelity $0.0760$ using $34$ evaluations of $F_{\mathrm{opt}}$, which in the fourth lowest basin with still very low infidelity.

\section{Discussion}\label{sec: sensitivity}
After seeing how time optimisation works in our examples. Let us recap how the optimised fidelity $F_{\mathrm{opt}}(T)$ varies with the evolution time $T$ and further discuss three reasons why the time optimisation is essential: 1) the presence of noise; 2) the oscillation of infidelity due to the drift Hamiltonian and 3) the ability to escape local traps in the other optimisation direction.

\emph{1. } At $T$ smaller than the minimal required time set by the quantum speed limit, we will expect the optimised fidelity $F_{\mathrm{opt}}(T)$ to increase as we increase $T$ since we have not yet had sufficient time to evolve to the target state at this point, due to the energy scale and constraints we place on our control Hamiltonian. At large $T$, the length of the evolution time is no longer the rate-limiting factor, and decay in $F_{\mathrm{opt}}(T)$ due to noise will dominate. Both of these effects can be seen in our experiments in 
\cref{fig:cost_func_bell_state,fig:cost_func_lmg}, and time optimisation is essential for identifying the optimal trade-off point between them. 

\emph{2. } If the control field cannot compensate for the effect of the drift field by some appropriate control pulses, then the control field will have limited influence on the trajectory in the state space that purely due to the drift field, which is some rotation along a hyper-surface. Such a rotation will periodically approach the target state, then move away and repeat, leading to oscillations in the optimised fidelity $F_{\mathrm{opt}}(T)$. Whether such oscillatory behaviour exists or not is not affected by the presence of noise, thus one can check whether $F_{\mathrm{opt}}(T)$ is oscillatory or not by simply performing the noiseless optimisation. We also can see that such an oscillation indeed comes from the fact that the control field cannot compensate for the effect of the drift field through the numerical experiment in \cref{appendix:identity_test}, in which we test whether the effective identity channel is achievable at different evolution times. In the presence of such oscillation, the optimised fidelity varies so significantly with time that time optimisation becomes essential. In some specific cases, we might be able to guess the position of the fidelity peak, but this cannot be done in general, especially in the presence of decoherence. We have discussed a more explicit derivation of this oscillation behaviour in \cref{sec:oscillation}. This is for the case when the drift Hamiltonian commutes with the control Hamiltonian and the basis for the drift Hamiltonian is not part of the basis for the control Hamiltonian, which is what happens in \cref{fig:cost_func_swap_1.0_M_8,fig:cost_func_dipole_0.5_M_8}.

\begin{figure}[htbp]
    \centering
    \includegraphics[width=0.495\textwidth]{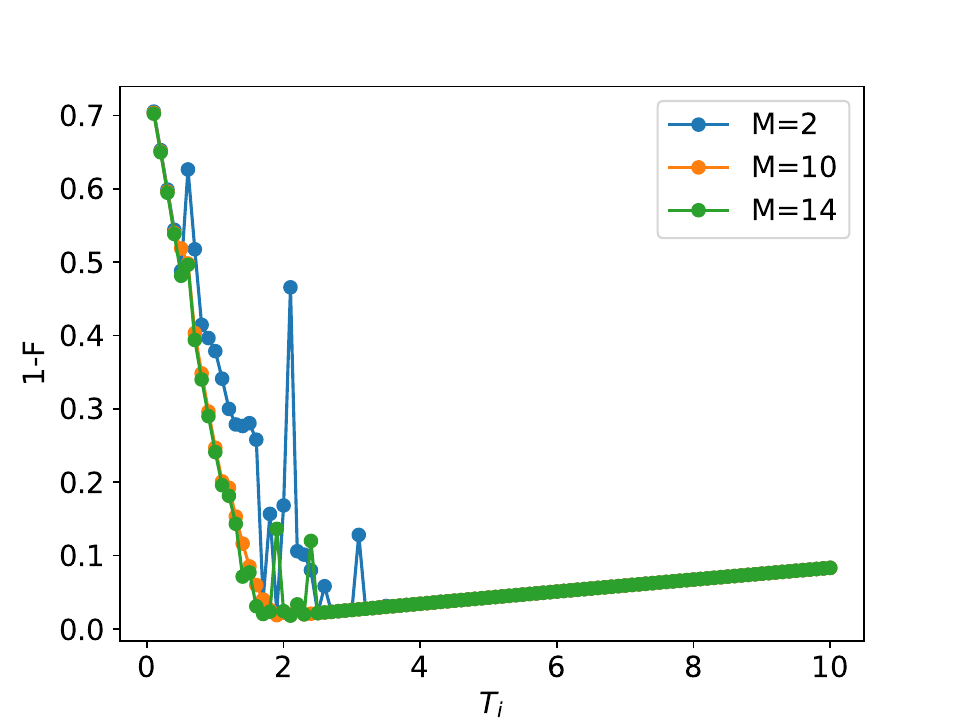}
    \caption{Optimised infidelity using basin-hopping for the state-to-state transfer problem of the LMG model (\cref{subsubsec: results:state-to-state-transfer: LMG}). $M$ is the number of basis functions in the pulse. Local traps in optimisation have led to fluctuations in the optimised infidelity. Such fluctuation is more prominent when $M$ is either too small (e.g. $M=2$) or too large (e.g. $M= 14$).}\label{fig:sensitivity_to_time_trade_off}
\end{figure}

\emph{3. } Variation in time can also help with escaping local traps in the pulse optimisation. When performing standard CRAB without $T$ optimisation, we are susceptible to local traps in the pulse optimisation in two main ways. When the number of basis functions $M$ is small, such local traps are due to the limited expressivity of the control pulse. When the number of basis function $M$ is big, such local traps are due to the difficulties in optimising $\vec{\alpha}$ due to the increased dimensionality. Both of these effects are shown in \cref{fig:sensitivity_to_time_trade_off}. There we see that we often can move away from local traps and reach lower infidelity by moving to another nearby $T$. Hence, by adding the evolution time as an additional parameter in the search space, the optimisation is more likely to navigate out of these false traps by moving along the new time direction. This is also seen in \cref{fig:cost_func_swap_1.0_M_8}. There, the cost function landscape obtained by CRAB is ragged near the second trough due to its inability to escape local traps. In contrast, basin-hopping is able to escape these local traps, reaching a lower infidelity than CRAB even at the same evolution time and giving us the true global optimum. 

\section{Conclusion}\label{sec:concl}
In this paper, we analyse the condition required for the noise to commute with the gate Hamiltonian in the context of quantum optimal control, which allow us to study the effect of such noise and obtain an analytic expression of the resultant fidelity. Under such noise, for a given evolution time, we can now perform the pulse optimisation using the CRAB protocol of the noisy system at a similar computation cost as the noiseless system, which is an exponential reduction in the computation cost in terms of the number of qubits. Leveraging this approach, we are able to perform optimisation of the evolution time on top of optimising the pulse parameters. We have performed numerical simulations on state-to-state transfer problems for Josephson charge qubits and LMG models and gate compilation problems for silicon spin qubits, under noise models such as global depolarising noise, local dephasing noise and dipole-dipole noise. In these examples, we indeed see a strong dependence of the optimised infidelity on the evolution time, caused by noise, drift field oscillations and local traps encountered in pulse optimisations. Our results indicate that an inappropriate choice of evolution time can significantly increase infidelity, highlighting the necessity to optimise the evolution time. Using the basin-hopping algorithm for optimisation, we are able to consistently identify globally optimal evolution times across all considered examples. In addition, we have explored the use of root-finding methods like bisection search, which can output a local optimum rather than a global one. However, these local optimum are nonetheless much better than an arbitrary choice of evolution time and is comparable to the global optimum in terms of infidelity in our examples.  

Our paper just marks the start of the numerous possibilities for incorporating time optimisation into quantum optimal control. A natural extension is to expand time optimisation to dCRAB~\cite{Rach_2015}, and more generally other quantum optimal control algorithms like GRAPE and Krotov method, to see if similar efficient implementation can be found. It is also interesting to explore the effect of more general noise models, for example going beyond Pauli noise or considering noise that only approximately commutes with the gate Hamiltonian. We can even consider pulse optimisation that incorporates Pauli-twirling-like behaviour that can enhance the commutation between the noise and the gate Hamiltonian. 

Another possible area to explore is the optimisation algorithm used. We have considered basin-hopping and the bisection method in this article, and one might wonder whether there are other optimisation algorithms that are more efficient and/or more accurate. Methods like simulated annealing~\cite{Vladim_1985(Simulated_Annealing)} and evolutionary methods~\cite{Vikhar_2016(DE)} have found previous success in quantum optimal control~\cite{He_2022(SA_1), Zhou_2020(SA_2), Qiu_2010(SDC_protocol), Bhole_2016(DE_1), Zahedinejad_2014(DE_2), Khurana_2017(DE_3), Ma_2015(DE_4)}, and thus will be interesting to investigate their performance with time optimisation. This can be new optimisers that more explicitly consider the difference between the cost function landscapes along the $T$ direction and the $\vec{\alpha}$ direction, or optimisers that can take advantage of the analytical expression of the fidelity expression that we derived.

\section*{Acknowledgements}
The authors would like to acknowledge the use of the University of Oxford Advanced Research Computing (ARC) facility in carrying out this work and specifically the facilities made available from the EPSRC QCS Hub grant (agreement No. EP/T001062/1). The authors also acknowledge support from EPSRC projects Robust and Reliable Quantum Computing (RoaRQ, EP/W032635/1) and Software Enabling Early Quantum Advantage (SEEQA, EP/Y004655/1).

\appendix
\section{Commutation between unitary and dissipative part}\label{sec:commute_noise}
The Liouville form of the Lindblad master's equation is 
\begin{align*}
    \dv{t} \pket{\rho} = \mathcal{L} \pket{\rho} = \left(\mathcal{L}_H + \mathcal{L}_D\right) \pket{\rho}
\end{align*}
where
\begin{align*}
\mathcal{L}_H &= -i (I \otimes H - H \otimes I)\\
\mathcal{L}_D &= \sum_{k} \left(L_k^* \otimes L_k - \frac{1}{2} \left(I \otimes L_k^\dagger L_k\right) - \frac{1}{2} \left((L_k^\dagger L_k)^* \otimes I\right)\right).
\end{align*}
Hence, the commutator between the unitary and dissipative part is
\begin{align*}
    &\left[\mathcal{L}_H, \mathcal{L}_D\right] = -i\sum_k \left\{ \left[I \otimes H,  L_k^* \otimes L_k\right] - \left[H \otimes I, L_k^* \otimes L_k\right]\right\} \\
    &\quad + \frac{i}{2}\sum_k \left\{\left[I \otimes H, I \otimes L_k^\dagger L_k\right]  -  \left[H \otimes I, (L_k^\dagger L_k)^* \otimes I\right] \right\}\\
    &= - i\sum_k \left\{ L_k^* \otimes \left[H, L_k\right] - \left[H, L_k^*\right]\otimes L_k \right\} \\
    &\quad + \frac{i}{2}\sum_k \left\{ I \otimes \left[H, L_k^\dagger L_k\right]  -  \left[H, (L_k^\dagger L_k)^*\right] \otimes I \right\}\\
\end{align*}
Let us define $C_k = \left[H, L_k\right]$, we then have:
\begin{align*}
    \left[H, L_k^*\right] &= \left[H, L_k\right]^* = C_k^*\\
    \left[H,  L_k^\dagger\right] &= - \left[H,  L_k\right]^\dagger = -C^\dagger\\
    \left[H, L_k^\dagger L_k\right] & = L_k^\dagger \left[H,  L_k\right] + \left[H, L_k^\dagger \right]L_k = L_k^\dagger C_k  - C_k^\dagger L_k \\
    \left[H, (L_k^\dagger L_k)^*\right] & = (L_k^\dagger C_k)^*  - (C_k^\dagger L_k)^*
\end{align*}
\begin{align*}
    &\left[\mathcal{L}_H, \mathcal{L}_D\right] = - i\sum_k \left\{ L_k^* \otimes C_k - C_k^*\otimes L_k \right\}  \\&+ \frac{i}{2}\sum_k \big\{ I \otimes \left(L_k^\dagger C_k  - C_k^\dagger L_k\right)  -  \left((L_k^\dagger C_k)^*  - (C_k^\dagger L_k)^*\right) \otimes I \big\}
\end{align*}
In this form, we can see that one possible way for this to be zero is to have
\begin{align}\label{eqn:commutation_condition_1}
    C_k = \left[H,  L_k\right] = \lambda_k L_k
\end{align}
for some real number $\lambda_k$, which can be verify by direct substitution. Physically this means that the jump operator $L_k$ will take an eigenvector $\ket{E}$ of $H$ with energy $E$ to another (unnormalised) eigenvector $L_k\ket{E}$ of energy $E+\lambda_k$:
\begin{align*}
    \left[H,  L_k\right] \ket{E} &= H L_k\ket{E} - L_k H\ket{E}  = \lambda_k L_k\ket{E}\\
    H L_k \ket{E} &= \left(E+\lambda_k\right) L_k \ket{E}
\end{align*}
Note that \cref{eqn:commutation_condition_1} also implies 
\begin{align*}
    \left[H, L_k^\dagger L_k\right] = L_k^\dagger C_k  - C_k^\dagger L_k = 0.
\end{align*}
This is a weaker condition than \cref{eqn:commutation_condition_1} and thus does not guarantee the commutation between $\mathcal{L}_D$ and $\mathcal{L}_H$. It ensures the forward plus backward jump preserves the eigenbasis of $H$, but not necessarily for individual jumps. Note that $L_k^\dagger L_k$ physically correspond to the decoherence rate of the $k$th decoherence process and its commutation with $H$ means it does not change with time. 

\section{Commutation and Fidelity of Pauli channel}\label{sec:pauli_channel}
\subsection{Pauli channel from Lindblad master equation}\label{sec:pauli_channel_Lindblad}
The set of Pauli operators is denoted as $\mathbb{G} = \{G_k\}_{k=0}^{4^N-1}$ with $G_0 = I$. For a given Pauli operator $G_k$, we can denote the correponding Pauli superoperator $\mathcal{G}_k$ acting on the incoming operator $\rho$ as $\mathcal{G}_k(\rho) = G_k\rho G_k^\dagger$. In this way, we can write the Lindblad master equation with Pauli jump operators $L_k = \sqrt{\gamma_k/2} G_k$ as:
\begin{align}\label{eqn:Lindblad_Pauli}
    \mathcal{L}_D & = \sum_{k=0}^{4^{N}-1}\frac{\gamma_k}{2} \left(\mathcal{G}_k - \mathcal{I}\right)
\end{align}
Do note that the contribution from the $k=0$ term is always $0$ since $\mathcal{G}_0 = \mathcal{I}$, thus we can set $\gamma_0$ to any number we want without affecting the dynamics.

In the rest of section, we will use the formalism of Pauli transfer matrix, which is essentially the matrix representation of the superoperator in the Pauli basis $\{2^{-N/2} \pket{G_k}\}$, where the factor of $2^{-N/2}$ is to normalise the Pauli basis such that $2^{-N}\pbraket{G_k}{G_k} = 1$. We will further use 
\begin{align*}
    \eta_{jk} = \eta(G_j, G_k) = G_{k} G_j G_{k}^{-1} G_{j}^{-1}
\end{align*}
to denote the commutator between $G_j$ and $G_k$. 

In this way, the action of $\mathcal{G}_k$ in the Pauli transfer matrix formalism is given by:
\begin{align}\label{eqn:pauli_to_ptm_1}
    \mathcal{G}_k \pket{G_j} = \begin{cases}
        \pket{G_j} \quad &\eta_{jk} = +1\\
        -\pket{G_j} \quad &\eta_{jk} = -1
    \end{cases} \nonumber\\
    \Rightarrow \quad \mathcal{G}_k = 2^{-N}\sum_{j=0}^{4^N-1} \eta_{jk} \pketbra{G_j}{G_j} 
\end{align}

Substituting back into \cref{eqn:Lindblad_Pauli}, we have:
\begin{equation}
    \begin{split}\label{eqn:Lindblad_dissip_pauli}
        \mathcal{L}_D & = 2^{-N} \sum_{j=0}^{4^N-1} \sum_{k=0}^{4^{N}-1}\frac{\gamma_k}{2}  (\eta_{jk} - 1) \pketbra{G_j}{G_j}\\
    & =  2^{-N} \sum_{j=0}^{4^N-1} \left(-\lambda_j\right) \pketbra{G_j}{G_j}
    \end{split}
\end{equation}
where
\begin{align}\label{eqn:decay_factor}
    \lambda_j = \sum_{k=0}^{4^{N}-1}(1 - \eta_{jk}) \frac{\gamma_k}{2} = \sum_{k,\ \eta_{jk} = -1}\gamma_k.
\end{align}
i.e. the dissipative Lindbladian is diagonalised in the Pauli basis, each associated with a decay constant $\lambda_j$ given by the sum of the strength of the individual noise components that anti-commute with $G_j$. Note that again the factor $2^{-N}$ is here to normalise the Pauli basis, i.e. the set of orthonormal basis is $\{2^{-N/2}\pket{G_j}\}$, it is not part of the eigenvalue. Since this is a diagonal matrix, it can be directly exponentiated to obtain the action of the resultant Pauli channel from the Lindbladian:
\begin{align}\label{eqn:Pauli_channel_action_PTM}
    e^{\mathcal{L}_D T} = 2^{-N} \sum_{j = 0}^{4^N - 1} e^{- \lambda_{j} T} \pketbra{G_j}{G_j}
\end{align}

In this way, we can calculate the fidelity between the noisy output state and the target state as:
\begin{align}\label{eqn:fid_decay}
    \pbra{\rho_{g}}e^{\mathcal{L}_D T}\pket{\rho_{f}} 
    = 2^{-N}\sum_{j=0}^{4^{N}-1} e^{- \lambda_{j} T} \pbraket{\rho_{g}}{G_j} \pbraket{G_j}{\rho_{f}} 
\end{align}
This is the extreme case in which all $\gamma_k$ are very different. In practice, there will be a lot of similar $\gamma_k$ and thus similar $\lambda_j$. The Pauli basis with the same $\lambda_j$ can be grouped together. 

Any unitary part $\mathcal{L}_H$ that is block diagonal in the same way as the degenerate subspaces of $\mathcal{L}_D$ will commute with $\mathcal{L}_D$ since $\mathcal{L}_D$ is proportional to identity in these subspaces. 

In another word, for $\mathcal{L}_{H}$ to commute with $\mathcal{L}_D$, for any given of Pauli basis $G_i$ and $G_j$ we require either $\lambda_i = \lambda_j$, or 
\begin{align*}
    &\bra{G_i}\mathcal{L}_{H}\pket{G_j} = \Tr(G_i\mathcal{L}_{H}(G_j)) \\
    &= -i \left(\Tr(G_iG_jH) - \Tr(G_jG_iH)\right)  = 0.
\end{align*}
A set of sufficient (but not necessary) conditions for the above equation to be true is
\begin{equation}\label{eqn:pauli_commute_cond}
    \begin{rcases}
        &\lambda_i = \lambda_j\\
        \quad \text{or} &\left[G_i, G_j\right] = 0\\
        \quad \text{or} &\left[G_i, H\right] = 0\\
        \quad \text{or} &\left[G_j, H\right] = 0\\
        \quad \text{or} &\Tr(G_iG_jH) = 0
    \end{rcases}
    \forall i, j \quad \Rrightarrow \quad \left[\mathcal{L}_H, \mathcal{L}_D\right] = 0
\end{equation}

\subsection{Transformation between Pauli transfer matrix and Pauli channels}
From the definition of the commutator between Pauli operators, we have:
\begin{align}\label{eqn:eta_ortho}
    \sum_{k=0}^{4^{N}-1} \eta_{ik}\eta_{jk} &= \sum_{k=0}^{4^{N}-1} \eta(G_i, G_k) \eta(G_j, G_k)\nonumber \\
    & = \sum_{k=0}^{4^{N}-1} \eta(G_iG_j, G_k)\nonumber\\
    & = 4^N \delta_{ij}
\end{align}
i.e. $2^{-N}\eta_{jk}$ is a orthogonal matrix, it is actually the $2N$ qubit Hadamard matrix with some column/row permutation. 

From \cref{eqn:pauli_to_ptm_1}, we know how to decompose a Pauli superoperator into the basis of the Pauli transfer matrix:
\begin{align}\label{eqn:pauli_to_ptm}
    \mathcal{G}_k = 2^{-N}\sum_{j=0}^{4^N-1} \eta_{jk} \pketbra{G_j}{G_j} 
\end{align}
Using \cref{eqn:eta_ortho}, we can also perform the reverse transformation:
\begin{align}\label{eqn:ptm_to_pauli}
    2^{-N}\sum_{k=0}^{4^{N}-1} \eta_{ik}\mathcal{G}_k & = \sum_{j=0}^{4^N-1}  \left(4^{-N}\sum_{k=0}^{4^{N}-1}\eta_{ik}\eta_{jk}\right) \pketbra{G_j}{G_j}\nonumber\\
    &= \pketbra{G_i}{G_i}
\end{align}
i.e. the orthogonal matrix $2^{-N}\eta_{jk}$ can transform between the pauli transfer matrix basis $\{\pketbra{G_j}{G_j}\}$ and the standard Pauli channel basis $\{\mathcal{G}_k\}$ (or equivalently between $\{2^{-N}\pketbra{G_j}{G_j}\}$ and $\{2^{-N}\mathcal{G}_k\}$).

We can use this to rewrite the resultant Pauli channel from the master's equation in \ref{eqn:Pauli_channel_action_PTM} into the standard form:
\begin{align}\label{eqn:Pauli_channel_action}
    e^{\mathcal{L}_D T} &= 2^{-N} \sum_{j = 0}^{4^N - 1} e^{- \lambda_{j} T} \pketbra{G_j}{G_j}\nonumber\\
    &= 2^{-N} \sum_{j = 0}^{4^N - 1} e^{- \lambda_{j} T} \left(2^{-N}\sum_{k=0}^{4^{N}-1} \eta_{jk}\mathcal{G}_k\right)\nonumber\\
    & = 4^{-N} \sum_{k=0}^{4^{N}-1} \left(\sum_{j = 0}^{4^N - 1} \eta_{jk} e^{- \lambda_{j} T}\right)  \mathcal{G}_k
\end{align}
i.e. the error probability of the $k$th Pauli operator is
\begin{align}\label{eqn:Pauli_error_prob}
    p_k = 4^{-N} \sum_{j = 0}^{4^N - 1} \eta_{jk} e^{- \lambda_{j} T}
\end{align}

\subsection{Example: dephasing noise}\label{sec:dephasing_noise}

For single-qubit dephasing channels, we simply have $\gamma_Z = \gamma$ and $\gamma_I = \gamma_X = \gamma_Y = 0$, and $\gamma$ here is the dephasing rate we input into our numerical simulation. Using \cref{eqn:decay_factor}, we thus have $\lambda_I = \lambda_Z = 0$ and $\lambda_X = \lambda_Y = \gamma$. Following \cref{eqn:Pauli_channel_action_PTM}, we have the Pauli transfer matrix representation of the channel:
\begin{align}\label{eqn:single_qubit_dephase_1}
    e^{\mathcal{L}_D T} &=   \frac{1}{2} \left(\pketbra{I}{I} + \pketbra{Z}{Z}\right) + \frac{1}{2} e^{- \gamma T} \left(\pketbra{X}{X} + \pketbra{Y}{Y}\right)
\end{align}
Using \cref{eqn:Pauli_error_prob}, we have
\begin{align*}
    p_I &= 4^{-N} \left(1 + e^{-\gamma T} + e^{- \gamma T} + 1\right) = \frac{1 + e^{-\gamma T}}{2}\\
    p_X &= 4^{-N} \left(1 + e^{-\gamma T} - e^{- \gamma T} - 1\right) = 0\\
    p_Y &= 4^{-N} \left(1 - e^{-\gamma T} + e^{- \gamma T} - 1\right) = 0\\
    p_Z &= 4^{-N} \left(1 - e^{-\gamma T} - e^{- \gamma T} + 1\right) = \frac{1 - e^{-\gamma T}}{2}
\end{align*}
Thus the corresponding Pauli channel following \cref{eqn:Pauli_channel_action} is
\begin{align}\label{eqn:single_qubit_dephase_2}
    e^{\mathcal{L}_D T} &=   \frac{1 + e^{- \gamma T}}{2} \mathcal{I} + \frac{1 - e^{- \gamma T}}{2} \mathcal{Z}
\end{align}

When we have $N$ qubits with individual qubits undergoing dephasing noise, the jump operators in the Master's equation are simply all single-qubit $Z$ operators with the coefficient $\sqrt{\gamma}$, and no other jump operators. Looking back at the gate Hamiltonian in \cref{eqn:rotating_frame_spin_spin_Hamiltonian}, we see that these jump operators commute with all the bases in the Hamiltonian, thus \cref{eqn:commutation_condition_1} is satisfied and we can study the unitary part and the noise part of the evolution separately. The Pauli transfer matrix of the resultant $N$-qubit channel from local dephasing is simply given as the tensor product of \cref{eqn:single_qubit_dephase_1}, which is
\begin{align*}
    e^{\mathcal{L}_D T} &= 2^{-N} \sum_{w = 0}^{N} e^{- w\gamma T} \sum_{j:\text{wt}_X(G_j) = w} \pketbra{G_j}{G_j}
\end{align*}
where $\text{wt}_X(G_j)$ is the weight of the $X$ string of $G_j$ in the symplectic representation, i.e. the number of qubits that is acted non-trivially by $X$ or $Y$. The corresponding stand form of the Pauli channel is given by the tensor product of \cref{eqn:single_qubit_dephase_2}. 

For example, for two qubits, we have its Pauli transfer matrix as:
\begin{align*}
     e^{\mathcal{L}_D T} 
    & = \frac{1}{4} \left(\pketbra{I}{I} + \pketbra{Z_1}{Z_1} + \pketbra{Z_2}{Z_2} + \pketbra{Z_1Z_2}{Z_1Z_2}\right)\\
    &+ \frac{e^{- \gamma T}}{4}  \big(\pketbra{X_1}{X_1}+ \pketbra{X_2}{X_2} +\pketbra{Y_1}{Y_1} + \pketbra{Y_2}{Y_2}\\
    &\quad \quad\quad \quad + \pketbra{Z_1X_2}{Z_1X_2} + \pketbra{X_1Z_2}{X_1Z_2}  \\
    &\quad \quad\quad \quad + \pketbra{Y_1Z_2}{Y_1Z_2} + \pketbra{Z_1Y_2}{Z_1Y_2}\big)\\
    &+ \frac{e^{- 2\gamma T}}{4}  \big(\pketbra{X_1X_2}{X_1X_2}+ \pketbra{Y_1Y_2}{Y_1Y_2} \\
    & \quad\quad\quad\quad + \pketbra{X_1Y_2}{X_1Y_2} + \pketbra{Y_1X_2}{Y_1X_2}\big)
\end{align*}

\subsection{Group channels}
A specific type of Pauli channel we want to discuss here is the group channel~\cite{caiMultiexponentialErrorExtrapolation2021}. Let $\widetilde{\mathbb{F}}$ be a set of independent Pauli operators and $\mathbb{F} = \expval*{\widetilde{\mathbb{F}}}$ to be the group of Pauli operator generated by this set, where all operators and composition here are defined without the irrelevant phase factors (modulo phase). The maximal group channel for the group of Pauli operator $\mathbb{F}$ is defined as the channel in which all of the elements in the group happen with equal probability:
\begin{align*}
    \mathcal{J}_{\mathbb{F}} = \frac{1}{\abs{\mathbb{F}}}\sum_{F_k \in \mathbb{F}}  \mathcal{F}_k = \prod_{\widetilde{F}_k \in \widetilde{\mathbb{F}}}  \frac{1 + \widetilde{\mathcal{F}}_k}{2}.
\end{align*}
We can see that when this channels acts on the different Pauli operators, we have:
\begin{align}\label{eqn:group_channel_prop}
    \mathcal{J}_{\mathbb{F}} (G_j) = \begin{cases}
        G_j \quad &\text{$G_j$ commute with all elements in $\widetilde{\mathbb{F}}$}\\
        0 \quad &\text{Otherwise}
    \end{cases}
\end{align}
Equivalently, we can also write it in the Pauli transfer matrix form as:
\begin{align}\label{eqn:group_channel_PTM}
    \mathcal{J}_{\mathbb{F}}  = 2^{-N} \sum_{G_j \in \mathbb{G}_{\mathbb{F}, +}} \pket{G_j} \pbra{G_j}
\end{align}
where $\mathbb{G}_{\mathbb{F}, +}$ is the set of Pauli operators that commute with all elements in $\widetilde{\mathbb{F}}$ (and thus $\mathbb{F}$).  This is actually a projection operator onto the subspace spanned by $\mathbb{G}_{\mathbb{F}, +}$. 

A general group channel of error probability $p$ simply means that there is probability $p$ that the maximal group error happens:
\begin{align*}
    \mathcal{J}_{\mathbb{F}, p} = (1-p) \mathcal{I}  + p \mathcal{J}_{\mathbb{F}} 
\end{align*}

Such group channels arise from the dissipative part of the master equation when the jump operators are $\sqrt{\frac{\gamma}{\abs{\mathbb{F}}}} F_k$ for all elements in the group $\mathbb{F}$:
Hence, in the superoperator form we have:

\begin{align*}
    \mathcal{L}_D & = \frac{\gamma}{\abs{\mathbb{F}}} \sum_{F_k \in \mathbb{F}}  \left(\mathcal{F} - \mathcal{I}\right) \\
    & = - 2^{-N} \gamma \sum_{G_j \not\in \mathbb{G}_{\mathbb{F}, +}} \pket{G_j} \pbra{G_j}
\end{align*}
Compared to \cref{eqn:Lindblad_dissip_pauli}, we see that this means
\begin{align}\label{eqn:decay_factor_group}
    \lambda_j = \begin{cases}
        0 \quad G_j \in \mathbb{G}_{\mathbb{F}, +}\\
        \gamma \quad G_j \not\in \mathbb{G}_{\mathbb{F}, +}
    \end{cases}
\end{align}
Hence, using \cref{eqn:Pauli_channel_action_PTM}, the resultant noise channel from the dissipator after time $T$ is given as:
\begin{equation}
    \begin{split}\label{eqn:evolution_op_group}
    &\quad    e^{\mathcal{L}_D T} \\
    &= 2^{-N} e^{-\gamma T}\sum_{G_j \not \in \mathbb{G}_{\mathbb{F}, +}} \pketbra{G_j}{G_j} + 2^{-N} \sum_{G_j \in \mathbb{G}_{\mathbb{F}, +}} \pketbra{G_j}{G_j}\\
    & = 2^{-N} e^{-\gamma T} \sum_{j} \pketbra{G_j}{G_j} + 2^{-N} (1-e^{-\gamma T}) \sum_{G_j \in \mathbb{G}_{\mathbb{F}, +}} \pketbra{G_j}{G_j}\\
    & = e^{-\gamma T} \mathcal{I} + (1-e^{-\gamma T})\mathcal{J}_{k}
    \end{split}
\end{equation}
i.e. this is a group channel with the maximal group error $\mathcal{J}_{k}$ occurring with the probability $(1-e^{-\gamma T})$.

The fidelity between the noisy output state and the target state is:
\begin{equation}
    \begin{split}\label{eqn:fid_decay_group}
    &\quad    \pbra{\rho_{g}}e^{\mathcal{L}_D T}\pket{\rho_{f}} \\
    & = 2^{-N} e^{-\gamma T}\sum_{G_j \not \in \mathbb{G}_{\mathbb{F}, +}} \pbraket{\rho_{g}}{G_j}\pbraket{G_j}{\rho_{f}} \\
    &\quad + 2^{-N} \sum_{G_j \in \mathbb{G}_{\mathbb{F}, +}} \pbraket{\rho_{g}}{G_j}\pbraket{G_j}{\rho_{f}} \\
    & = e^{-\gamma T} \pbraket{\rho_{g}}{\rho_{f}} + 2^{-N}(1-e^{-\gamma T})\sum_{G_j \in \mathbb{G}_{\mathbb{F}, +}} \pbraket{\rho_{g}}{G_j}\pbraket{G_j}{\rho_{f}}
    \end{split}
\end{equation}

\subsection{Example: Depolarising channel}
\label{appendix:subsec:depolarising_channel}
For global depolarising channels, the noise group being the entire Pauli group $\mathbb{F} = \mathbb{G}$, thus the commuting basis consists of only the identity operator: $\mathbb{G}_{\mathbb{F}, +} = \{I\}$. 

Hence, using \cref{eqn:decay_factor_group}, we have
\begin{align*}
    \lambda_0 &= 0\\
    \lambda_j &= \gamma \quad \forall j \neq 0
\end{align*}

Looking back at \cref{eqn:pauli_commute_cond}, we have:
\begin{align*}
    \begin{rcases}
        i = 0 \text{ or } j = 0 \quad &\Rightarrow \left[G_i, G_j\right] = 0\\
        i \neq 0  \text{ and } j = 0 \quad &\Rightarrow \lambda_i = \lambda_j\\
    \end{rcases}
     \Rrightarrow \quad \left[\mathcal{L}_H, \mathcal{L}_D\right] = 0
\end{align*}
for any $\mathcal{L}_H$. Thus, the depolarising channel commutes with all unitary parts of the master equation and the resultant fidelity following \cref{eqn:fid_decay_group} is given by:
\begin{align*}
    &\quad \pbra{\rho_{g}}e^{\mathcal{L}_D T}\pket{\rho_{f}} \\
    &= e^{-\gamma T} \pbraket{\rho_{g}}{\rho_{f}} + 2^{-N}(1-e^{-\gamma T}) \pbraket{\rho_{g}}{I}\pbraket{I}{\rho_{f}} \\
    &= e^{-\gamma T} \pbraket{\rho_{g}}{\rho_{f}} + 2^{-N}(1-e^{-\gamma T}).
\end{align*}

\subsection{Example: Two-qubit Dipole-Dipole Noise Channel}
\label{appendix:subsec:dipole_dipole_channel}
By dipole-dipole channel, we mean the Pauli channel with the noise group 
\begin{align}
    \mathbb{F} = \{I, Z_1Z_2\}.
\end{align}
which leads to the jump operators:
\begin{align*}
    L_{0} &= \sqrt{\frac{\gamma}{2}}I \\
    L_{1} &= \sqrt{\frac{\gamma}{2}}Z_{1} Z_{2}
\end{align*}
Here $\gamma$ is the decay rate. 

Looking back at the gate Hamiltonian in \cref{eqn:rotating_frame_spin_spin_Hamiltonian_first_order_approx}, we see that these jump operators commute with all the basis in the Hamiltonian, thus \cref{eqn:commutation_condition_1} is satisfied and we can study the unitary part and the noise part of the evolution separately. This means the resultant fidelity follows \cref{eqn:fid_decay_group}. All we need to do is to obtain $\mathbb{G}_{\mathbb{F},+}$, which is the Pauli operators that commute with the noise group $\mathbb{F}$. It consists of all Pauli operators that have even weights in the $X$ part of the symplectic representation, which is generated by
\begin{align}
    \widetilde{\mathbb{G}}_{\mathbb{F},+} = \{Z_{1}, Z_{2}, X_{1}X_{2}\}
\end{align}

\subsection{The Application of Noise channels for Gate Compilation} \label{appendix:subsec:noise_channel_gate_compilation}

When we use the scheme using the Choi state of the channel as noted in \cref{subsubsec:choi_state}, we need to modify the definition of the noise channels in \cref{appendix:subsec:dipole_dipole_channel} and \cref{sec:dephasing_noise} because the number of qubits of the Choi state is twice the size of the number of qubits the gates are acted upon. Thus, for example for two-qubit gate compilations, the Choi state will be a four-qubit state.

Since the error channel and the unitary operation are performed on the two original qubits before the bending of the quantum circuit in \cref{fig:choi_state}, the operations should be acted on either all odd-numbered qubits or all even-numbered qubits of the Choi state. In this paper, we chose the convention of performing operations on all odd-numbered qubits. The target state would be the same except we perform the target gate on all odd-numbered qubits.

This changes the error channels to be four-qubit channels instead of the original two-qubit channels with the identity operators included in between the original operations. For example, the dipole-dipole error channel would be modified to a four-qubit channel as below:

\begin{align} \label{eqn:dipole_dipole_noise_four_qubits}
    \mathcal{Z}_{DD}(\rho) &= (1- \frac{p}{2})I^{\otimes 4}\rho I^{\otimes 4} \nonumber \\ 
    &+ \frac{p}{2}(Z_{1} \otimes I \otimes Z_{3} \otimes I)\rho(Z_{1} \otimes I \otimes Z_{3} \otimes I).
\end{align}The evolution of the density matrix would still follow the general arguments given in \cref{sec:commute_noise} and \cref{sec:pauli_channel}.

\section{Choi States}\label{subsubsec:choi_state}

\begin{figure}
    \centering
    \includegraphics[width=\linewidth, trim={3.5cm 4cm 3.5cm 4cm},clip]{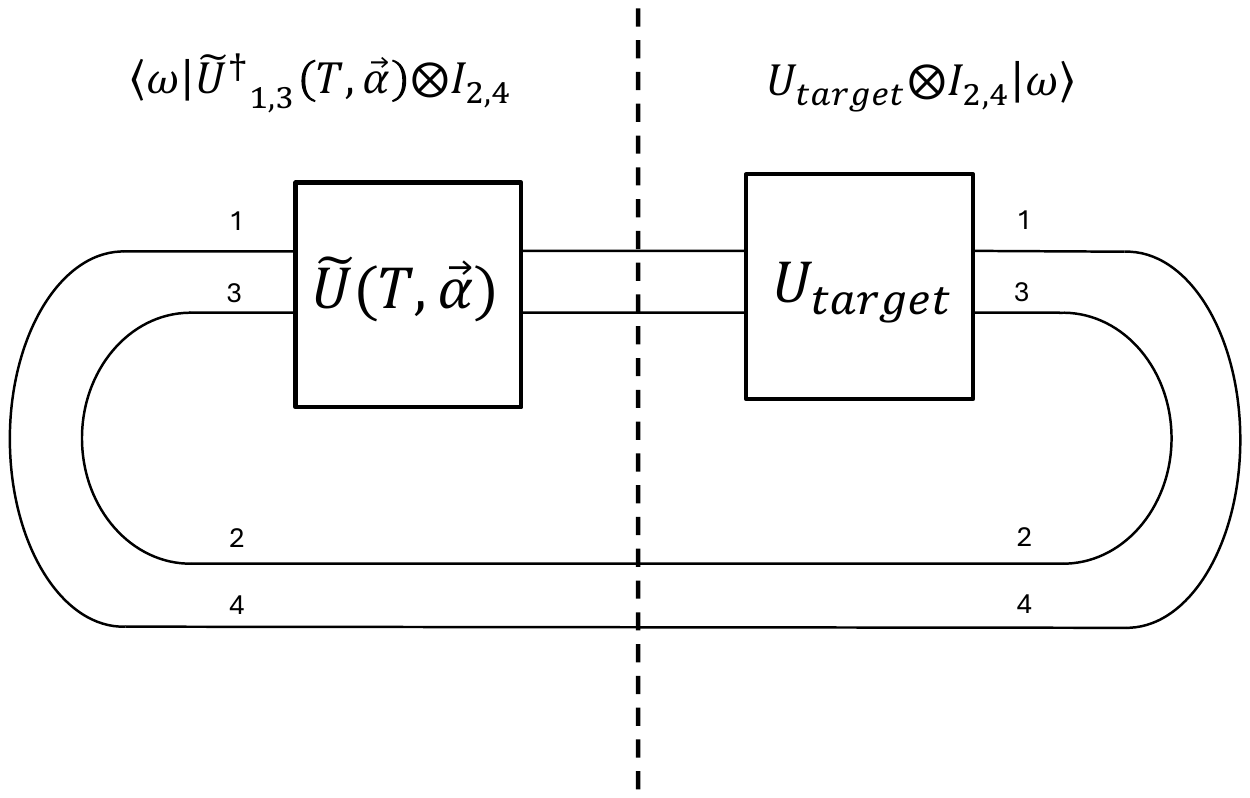}
    \caption{The tensor network diagram to compare the gate fidelity of two 2-qubit gates, i.e. $\tilde{U}(T, \vec{\alpha})$ and $U_{target}$, using the Choi state $U_{target}\ket{\omega}$. For 2-qubit gates, $\ket{\omega}$, is two Bell pairs, i.e. $\ket{\Psi^{+}} \otimes \ket{\Psi^{+}}$. The subscripts of the gates denote the qubits that the gate is acting on.}
    \label{fig:choi_state}
\end{figure}

The Choi-Jamio\'{l}kowski isomorphism tells us that, for any completely positive trace-preserving map, $\mathcal{E}$, there is a corresponding Choi state, $(\mathcal{E} \otimes \mathcal{I})(\ketbra{\omega}{\omega})$, where $\ket{\omega}$ are Bell pairs. When the map $\mathcal{E}$ is a unitary channel $\mathcal{E}(\rho) = U \rho U^{\dag}$, the Choi state becomes a pure state $(U \otimes I)\ket{\omega}$. We used this to map the gate compilation problems to state-to-state transfer problems, such that the initial state is the Bell pairs and the final state is the Choi state of the target gate, i.e. the CZ gate. See \cref{fig:choi_state} to see the tensor diagram representation of this scheme. The gate fidelity between two unitary operations, $\tilde{U}(T, \Vec{\alpha})$ and $U_{target}$, is equivalent to the state fidelity between $\tilde{U}\ket{\omega}$ and $U\ket{\omega}$:

\begin{align*}
    \frac{1}{2^{N}}Tr(\tilde{U}^{\dag}(T, \Vec{\alpha})U_{target}) \\
    =  \bra{\omega}(\tilde{U}^{\dag}(T, \Vec{\alpha})\otimes I)(U_{target}\otimes I)\ket{\omega},
\end{align*}
where $N$ is the number of qubits. Since we assume the error channels commuting with the Hamiltonian, we can use the results of \cref{sec:noisy_crab} to obtain the gate fidelity of $\tilde{U}(T, \vec{\alpha})$ and $U_{target}$ subject to error channels (See \cref{appendix:subsec:noise_channel_gate_compilation} for more details).

\section{Numerical Simulations}\label{sec:numerical_simulations}

\subsection{Implementation of CRAB and TCRAB}\label{sec:implementation_details}
The numerical integration in \cref{eqn:control_unitary_continuous} is performed by first-order Trotterisation with time step size $\Delta t$. 

\begin{figure}[h]
    \centering
    \includegraphics[width=\linewidth, trim={0 0 0 1cm},clip]{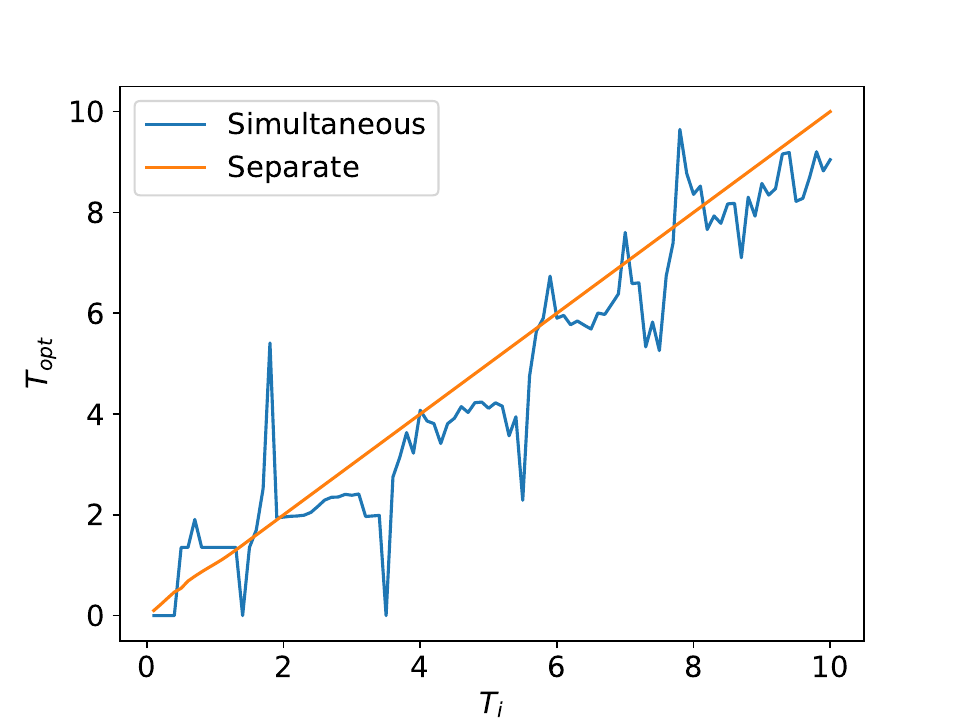}
    \caption{The optimal time, $T_{opt}$, obtained by the TCRAB algorithm with simultaneous (blue line) and separate (orange line) optimisation of $T$ and $\Vec{\alpha}$, for varying initial guesses of evolution time, $T_{i}$. The simultaneous optimisation yields more variation in optimal time than the separate optimisation, and it is more likely to be stuck at the local minima for the separate optimisation. The optimisations were performed for entanglement generation of two capacitively coupled Josephson charge qubits (See \cref{subsubsec: results:state-to-state-transfer: JosephsonChargeQubits}.). Note that we used a local optimiser, L-BFGS-B, with $2000$ as its maximum number of function evaluations.}
\label{fig:T_op_vs_T_init_sep_vs_sim}
\end{figure}

In TCRAB, we optimise $F(T, \vec{\alpha})$ over both $T$ and $\vec{\alpha}$. We employed two different optimisation methods: Basin-hopping with L-BFGS-B as its local optimiser and bisection method. In basin-hopping, we optimise $T$ and $\Vec{\alpha}$ simultaneously as we observed that optimising $\Vec{ \alpha}$ first and optimising $T$ later resulted in a local minimum that cannot be escaped for the temporal optimisation. We performed both simultaneous and separate optimisation of $T$ and $\vec{\alpha}$ for the entanglement generation of two capacitively coupled Josephson charge qubits (See \cref{subsubsec: results:state-to-state-transfer: JosephsonChargeQubits}). \cref{fig:T_op_vs_T_init_sep_vs_sim} shows the optimal evolution time obtained by L-BFGS-B from the simultaneous and separate optimisation of parameters, $\vec{\alpha}$ and $T$, with different initial guesses, $T_{i}$. For the separate optimisation, the maximum change of evolution time from its initial value was $0.085$, and changes of evolution time were in the order of $10^{-3}$ or below when $T_{i}$ was bigger than $1.0$. In contrast, for the simultaneous optimisation, the change of evolution time was more drastic so the maximum change of evolution time was around $3.61$.

Starting from an initial search interval, the bisection method finds the root of a function by iteratively narrowing down the search interval. The function we optimise is the derivative of the optimised infidelity by the evolution time, i.e. $F_{opt}(T)$ in \cref{eqn:crab_infid}. Like in basin-hopping, we used L-BFGS-B for local optimisation at each evolution time when evaluating $F_{opt}(T)$. We used the first-order finite difference approximation to estimate the derivative of $F_{opt}(T)$.

The goal of our simulations is to benchmark the ability of TCRAB to find the optimal parameters, i.e. $\vec{\alpha}_{opt}$, $T_{opt}$, at the global minimum of the infidelity. \cref{code:CRAB_TCRAB_benchmark} shows the pseudo-code of the benchmark. We take $N_{S}$ equal time slices in the range of possible evolution time, i.e. $[0, T_{max}]$: $\mathbb{T}_{\mathrm{init}} = \{T_{max}/N_{S}, 2T_{max}/N_{S}, ..., T_{max}\}$. $\mathbb{T}_{\mathrm{init}}$ is a set of initial evolution times for each run of CRAB and TCRAB. The frequencies of the truncated basis, $\{\omega\}_{m=1...M}$, were taken to be the same for each run of CRAB and TCRAB.

\begin{figure}
\begin{algorithm}[H]
	\caption{Benchmark of TCRAB} 
	\label{code:CRAB_TCRAB_benchmark}
    \begin{algorithmic}[1]
		\For {$i=1,2,\ldots, N_{S}$}
                \State Select the $i$th element of $\mathbb{T}_{\mathrm{init}}$ to be the time of evolution,  i.e. $T_{i} = T_{max}/N_{S} \times i$.
                \State Perform the CRAB with the evolution time $T_{i}$.
			\State Perform the TCRAB with the evolution time $T_{i}$ as the initial guess of optimal time in Basin-hopping.
		\EndFor
            \State Perform TCRAB using the bisection method until convergence.
            \State Among $N_{s}$ runs of TCRAB with different initial guesses of evolution times, $T_{i}$, the result with the lowest infidelity becomes the optimal time and the corresponding optimal pulse. Note down the occurrence of this optimal time in plot 3 of \cref{code:line:plots}. \label{code:line:TCRAB}
               \State Using the optimisation results of CRAB and TCRAB, generate three plots: 
               \begin{itemize}
                  \item \textbf{(Plot 1)}: Final infidelity after optimisation vs. initial evolution time, i.e. $T_{i}$, using $N_{s}$ runs of CRAB. In the same plot, draw optimal time and infidelity found by TCRAB with two optimisation methods, i.e. basin-hopping and bisection method.
                  \item \textbf{(Plot 2)}: The number of function evaluations vs. initial evolution time, $T_{i}$. (Only for CRAB and basin-hopping)
                  \item \textbf{(Plot 3)}: Histogram of final optimised time of the basin-hopping runs.
                \end{itemize} \label{code:line:plots}
            \State Refer to  plot 1 of \cref{code:line:plots} to compare the optima found by TCRAB with $N_{s}$ runs of CRAB. Check if the optimal time and infidelity found by TCRAB roughly match those of the CRAB run that resulted in the lowest infidelity.
	\end{algorithmic}
\end{algorithm}
\end{figure}

We first run CRAB optimisation on the problem of interest. In particular, we sweep the evolution time, i.e. $T_{i} \in \mathbb{T}_{\mathrm{init}}$ for each CRAB run. We can infer the optimal time by identifying the evolution time of the CRAB run that resulted in the lowest infidelity. In practice, we draw a plot of the infidelity against the evolution time, denoted as plot $1$ in \cref{code:line:plots}  of \cref{code:CRAB_TCRAB_benchmark}. Since time is not optimised for the runs of CRAB, the optimal time can be inferred from this plot by finding the evolution time where the final infidelity is the lowest. Note that the optimal time identified with CRAB runs is always an element in $\mathbb{T}_{\mathrm{init}}$, and it only serves the purpose of identifying the rough region where the true optimal time will be. The true optimal time will be inferred from TCRAB.

We run TCRAB optimisation on the same problem for both basin-hopping and the bisection method. The results of basin-hopping can vary due to the initial guess of optimal time. As we previously swept the evolution time of CRAB runs, we swept the initial guess of optimal evolution time, $T_{i} \in \mathbb{T}_{\mathrm{init}}$. Note that TCRAB optimises the evolution time, and it is the initial guesses of evolution time, but not the evolution times themselves, that are swept. Then, we identify the optimal evolution time by finding the evolution time of the basin-hopping run that resulted in the lowest infidelity. Furthermore, we check the fraction of basin-hopping runs that succeeded in obtaining the optimal time by looking at the histogram denoted as plot $3$ in the \cref{code:line:plots} of \cref{code:CRAB_TCRAB_benchmark}. 

We set the initial search interval of the bisection method to be $[0,T_{max}]$.  We used two stopping conditions: tolerance of the derivative and the length of the interval. If the derivative is smaller than a threshold or if the length of the search interval is smaller than a threshold, the algorithm converges. We compare the optimal time and infidelity with the results of CRAB and basin-hopping.

\subsection{Hyper-parameters}\label{subsec: hyper-parameters}

There is a set of hyper-parameters that the user has to specify to run either CRAB and TCRAB: The number of frequencies, $M$, the maximum frequency, $\omega_{max}$, and the set of basis frequencies $\Vec{\omega}$ selected based on these constraints. The number of frequencies determines the number of basis functions to express the pulse, which is $2 \times M$. For simulations in \cref{sec: results}, we chose to use $8$ basis functions, i.e. $M=8$ with an exception to the LMG model in \cref{subsubsec: results:state-to-state-transfer: LMG} where we chose $M=10$. For simulations in \cref{sec: sensitivity}, the number of frequencies was varied from $2$ to $14$. The maximum frequency is set to mimic realistic signal generators that are bandwidth-limited. Note that the frequencies, $\Vec{\omega}$ were drawn from a uniform distribution in $[0, \omega_{max}]$. In all simulations, we set the maximum frequency to be $20$, i.e. $\omega_{max}=20$.

There are additional hyper-parameters to run basinhopping\cite{Olson_2012(BasinHopping)} with L-BFGS-B\cite{Zhu_1997(L-BFGS-B)} as its local optimiser. Among many hyperparameters in the Scipy\cite{2020SciPy-NMeth}, we varied the following with the corresponding argument names in brackets: the maximum number of function evaluations (\textit{maxfun}), lower and upper bounds of the optimisation parameters to define the search space of $T$ and $\Vec{\alpha}$ (\textit{bounds}), the step size in numerical differentiation (\textit{eps}), the tolerance levels for the stopping criteria based on the values of the function and gradient (\textit{ftol}, \textit{gtol}).

The maximum number of function evaluations was set to be $10000$ for both CRAB and TCRAB. The search space of each component of $\Vec{\alpha}$ was bounded by $-100$ and $100$ such that $\Vec{\alpha} \in [-100, 100]^{\otimes M}$, for both CRAB and TCRAB, and the search space of $T$ was bounded by $[0, T_{max}]$, as noted in \cref{sec:TCRAB_theory}. We fixed the upper bound of the search space, $T_{max}$, to be $10$ for all simulations. The step size was chosen as $10^{-6}$. The tolerance levels, i.e. \textit{ftol} and \textit{gtol}, were $10^{-8}$ and $10^{-12}$. Other parameters were left as the default values of the implementation of L-BFGS-B in Scipy\cite{2020SciPy-NMeth}.

There are three additional hyper-parameters to run the bisection method: the step size in time used to evaluate the derivative of $F_{opt}(T)$, the tolerance level for two stopping conditions, i.e. the absolute value of $\dot{F}_{opt}(T)$ and the length of search interval. For the entanglement generation using Josephson charge qubits in \cref{subsubsec: results:state-to-state-transfer: LMG}, the step size and the two tolerance levels were $1\mathrm{e}{-3}$, $1\mathrm{e}{-6}$, $1\mathrm{e}{-6}$, respectively. For the rest of the problems, the step size and the two tolerance levels were $1\mathrm{e}{-4}$, $1\mathrm{e}{-6}$, $1\mathrm{e}{-4}$, respectively.

There are hyper-parameters of the state vector simulation. The number of time steps, $N_{t}$, was chosen to be $300$. Furthermore, we chose decay factors such that the decaying effect is visible in the search space. The step size of time chosen to sweep the search space was $0.1$ such that $\mathcal{I}_{100}= \{0.1, 0.2, ... 10\}$.

\section{Additional Numerics}\label{sec:add_numerics}
\begin{figure*}[htbp]
    \centering
    \subfloat[\label{fig:nfev_bell_state}]{\includegraphics[width=0.32\textwidth]{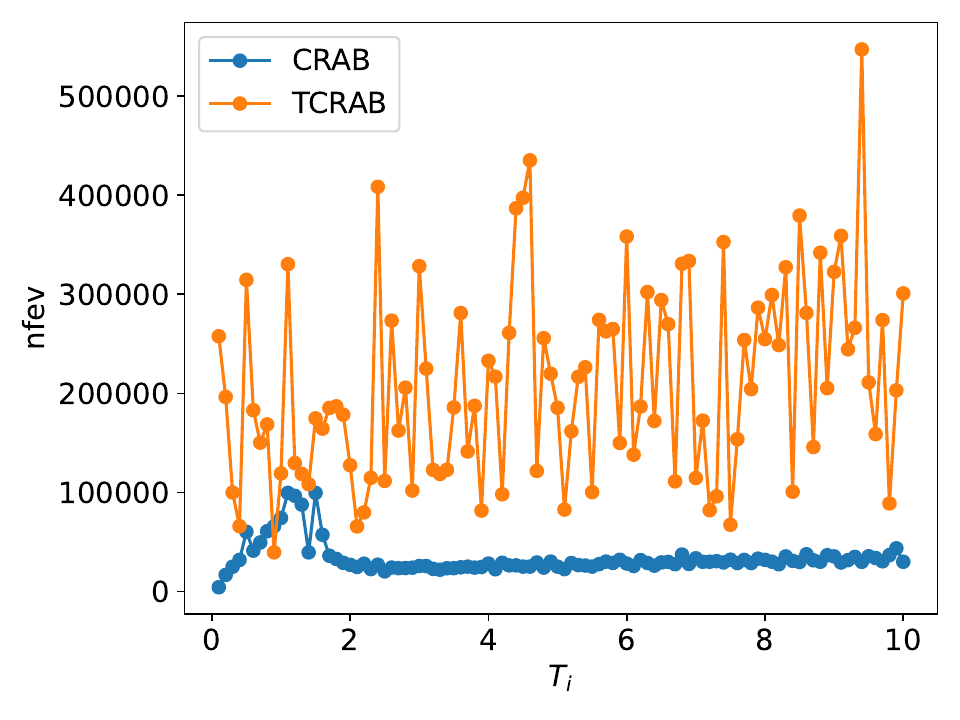}}
    \subfloat[\label{fig:optimal_T_bell_state}]{\includegraphics[width=0.32\textwidth]{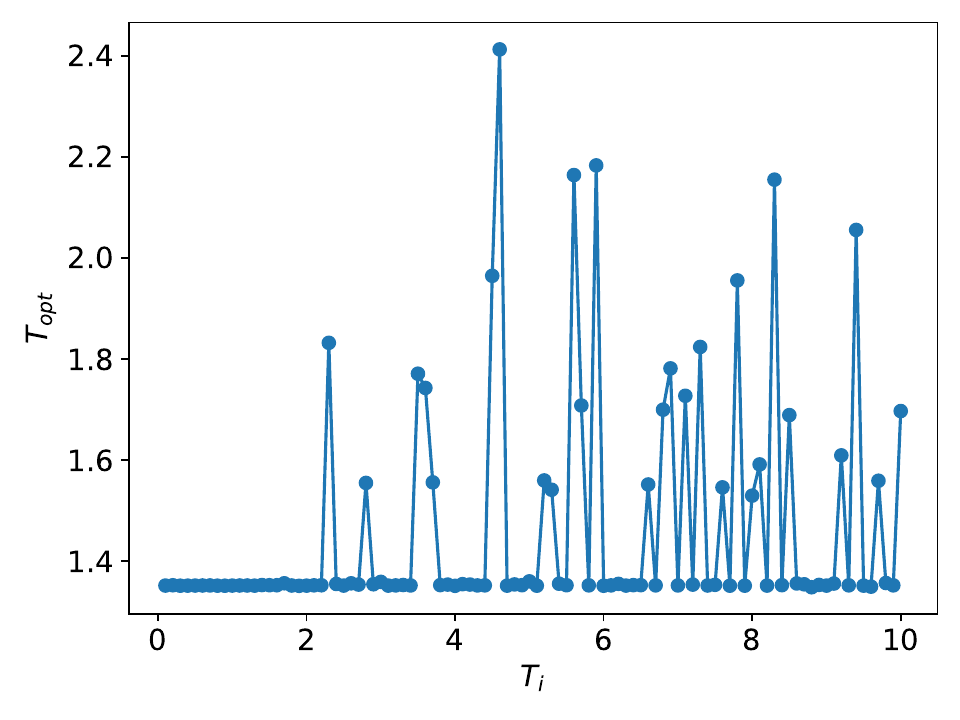}}
    \subfloat[\label{fig:optimal_T_hist_bell_state}]{\includegraphics[width=0.32\textwidth]{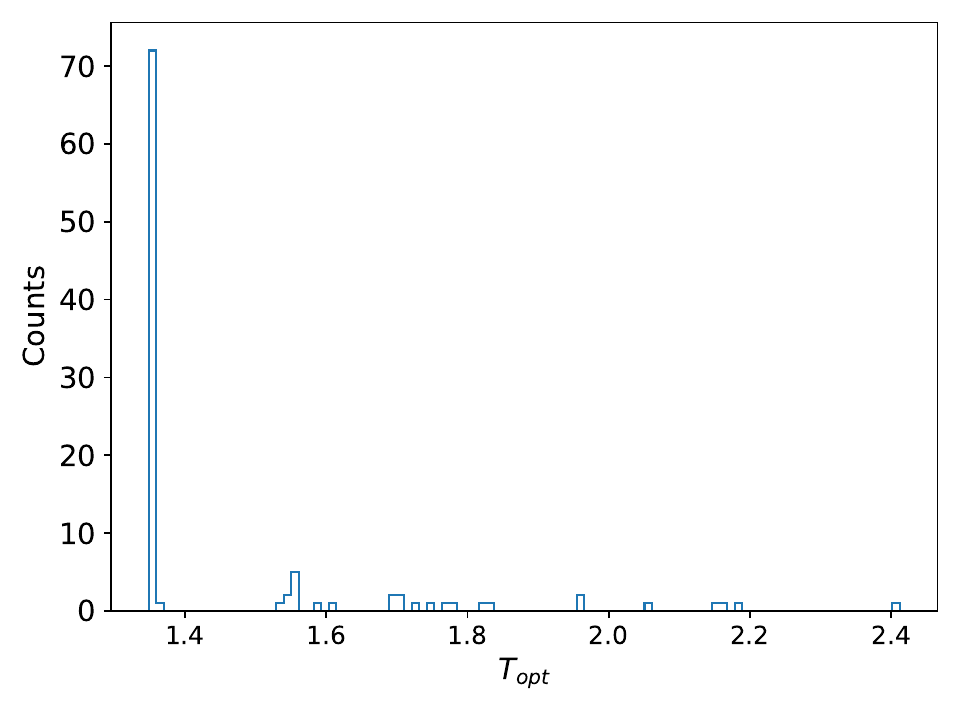}}

    \caption{Results of the entanglement generation of two capacitively coupled Josephson charge qubits are shown: (a) The number of function evaluations, $nfev$, to reach the convergence, and (b) The resulting optimal time, $T_{opt}$, for each initial evolution time (CRAB)/ initial guess of optimal time (TCRAB), $T_{i}$. (c) The distribution of optimal time, $T_{opt}$, found by the TCRAB scheme.}
\label{fig:bell_state_M_8}
\end{figure*}

\begin{figure*}[htbp]
    \centering

    \subfloat[\label{fig:nfev_lmg}]{\includegraphics[width=0.32\textwidth]{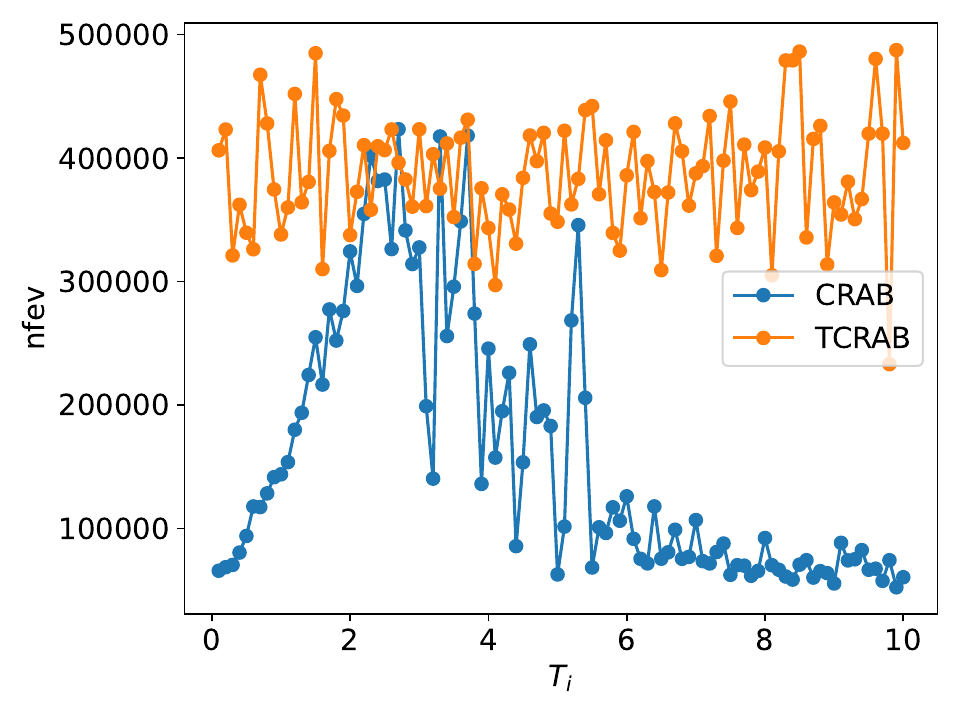}}
    \subfloat[\label{fig:optimal_T_lmg}]{\includegraphics[width=0.32\textwidth]{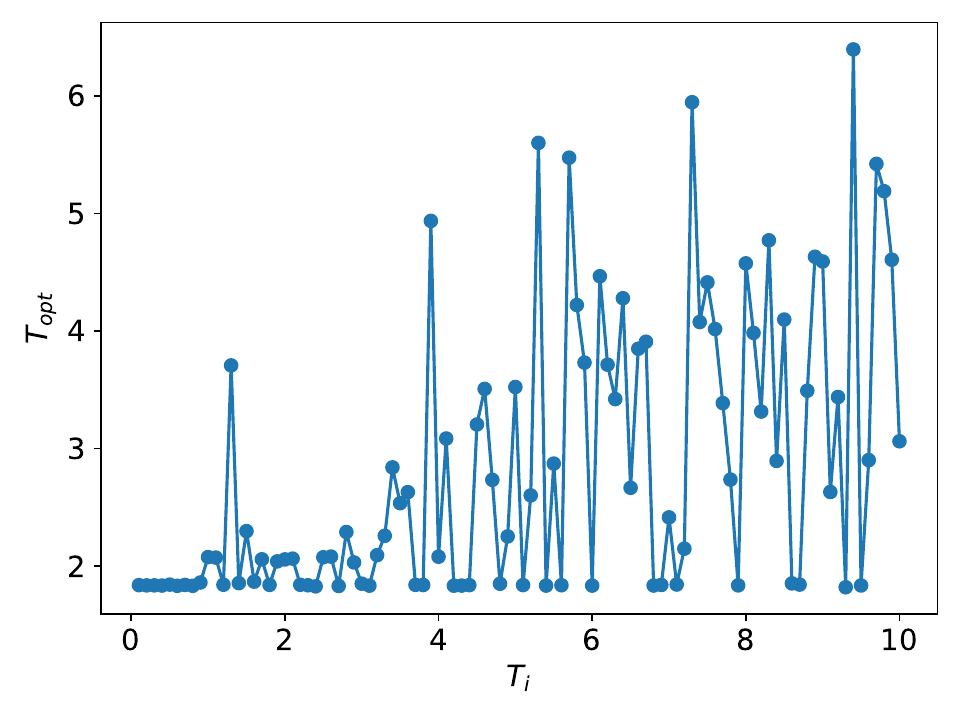}}
    \subfloat[\label{fig:optimal_T_hist_lmg}]{\includegraphics[width=0.32\textwidth]{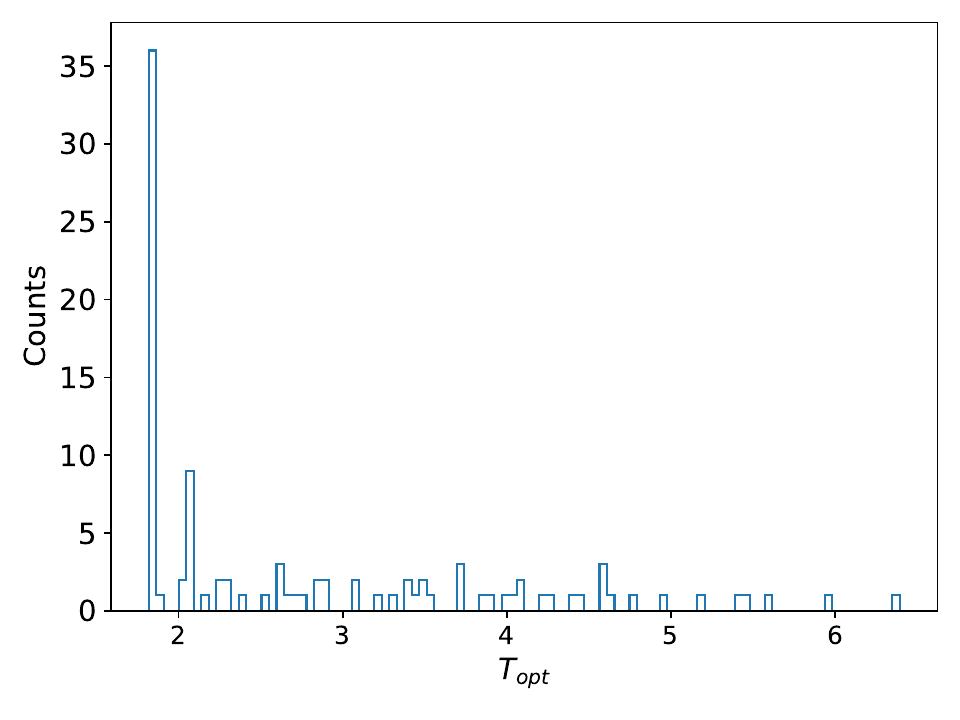}}

    \caption{Results of the state-to-state transfer from the ground state of paramagnetic phase to a ground state of ferromagnetic phase are shown: (a) The number of function evaluations, $nfev$, to reach the convergence, and (b) The resulting optimal time, $T_{opt}$, for each initial evolution time (CRAB)/ initial guess of optimal time (TCRAB), $T_{i}$. (c) The distribution of optimal time, $T_{opt}$, found by the TCRAB scheme.}
\label{fig:lmg_M_8}
\end{figure*}

\begin{figure*}[htbp]
    \centering
    \subfloat[\label{fig:nfev_swap_1.0_M_8}]{\includegraphics[width=0.32\textwidth]{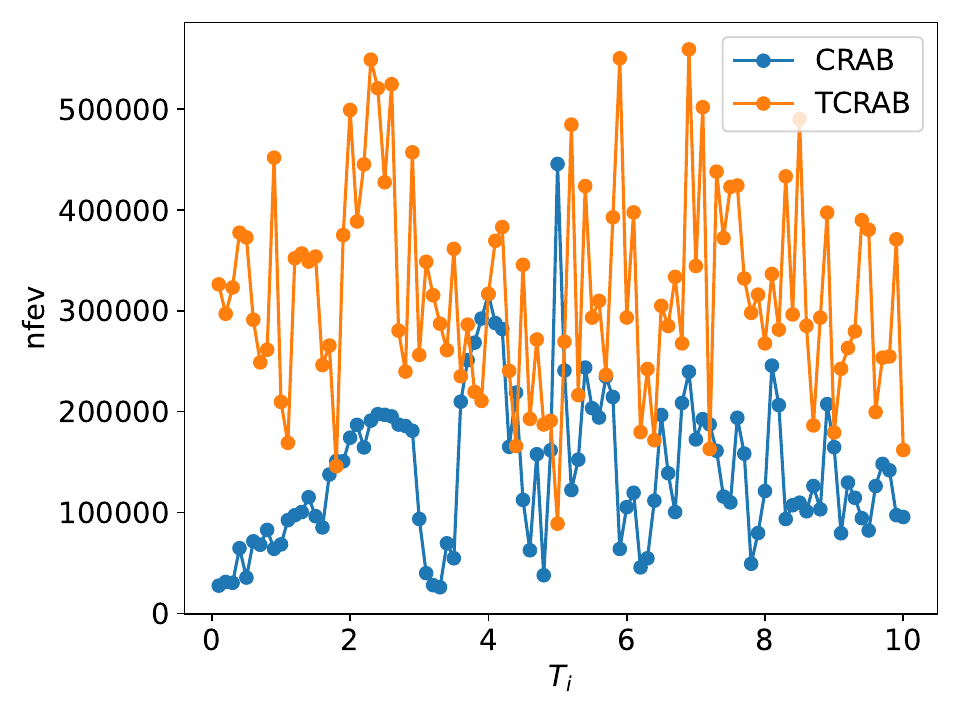}}
    \subfloat[\label{fig:optimal_T_swap_1.0_M_8}]{\includegraphics[width=0.32\textwidth]{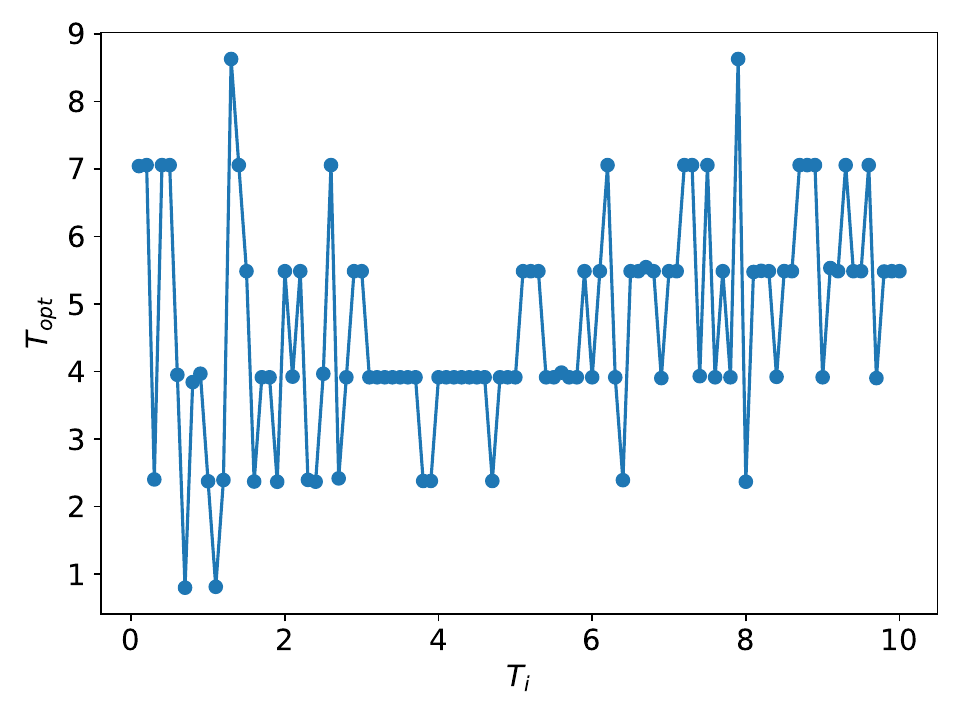}}
    \subfloat[\label{fig:optimal_T_hist_swap_1.0_M_8}]{\includegraphics[width=0.32\textwidth]{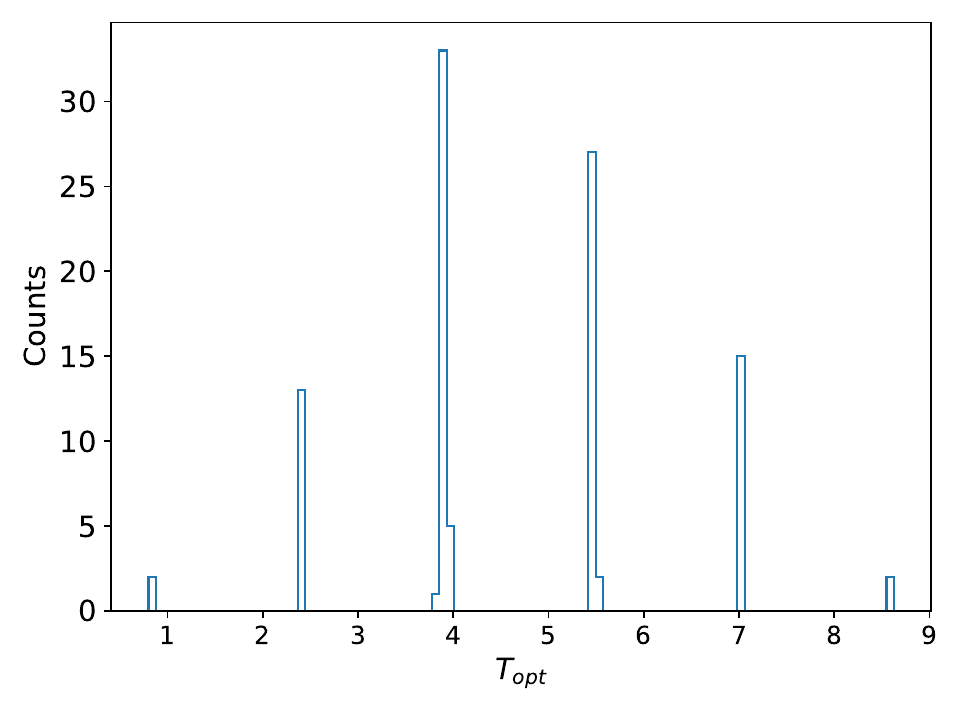}}
    \caption{Results of the gate compilation of CZ for $\Omega \ll J$ are shown: (a) The number of function evaluations, $nfev$, to reach the convergence, and (b) The resulting optimal time, $T_{opt}$, for each initial evolution time (CRAB)/ initial guess of optimal time (TCRAB), $T_{i}$. (c) The distribution of optimal time, $T_{opt}$, found by the TCRAB scheme.}
\label{fig:swap_1.0_M_8}
\end{figure*}

\begin{figure*}[htbp]
    \centering
    \subfloat[\label{fig:nfev_dipole_0.5_M_8}]{\includegraphics[width=0.32\textwidth]{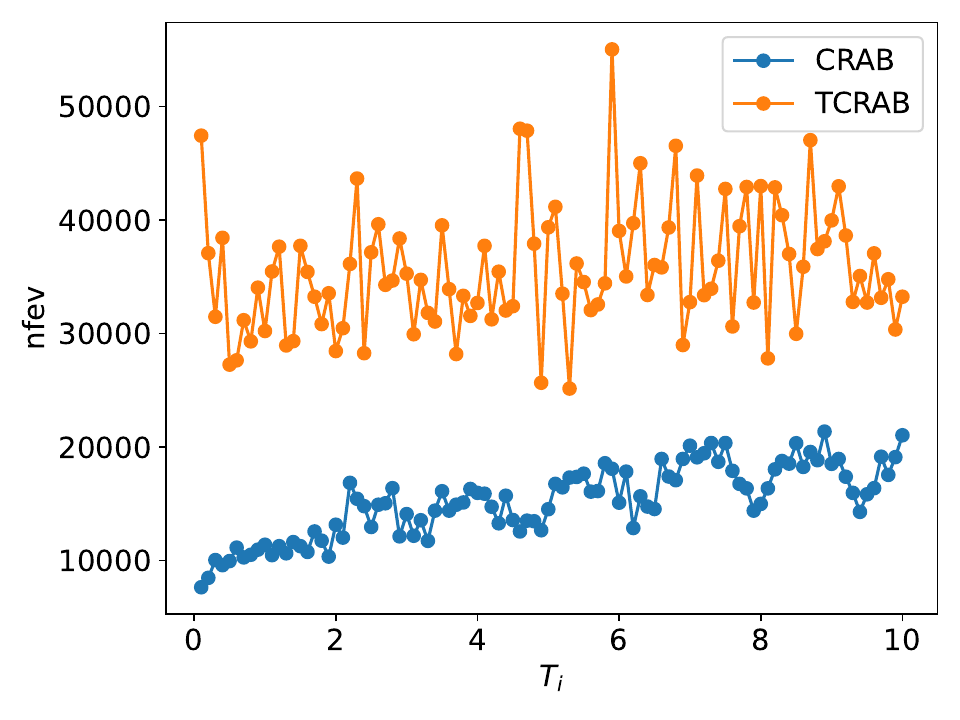}}
    \subfloat[\label{fig:optimal_T_dipole_0.5_M_8}]{\includegraphics[width=0.32\textwidth]{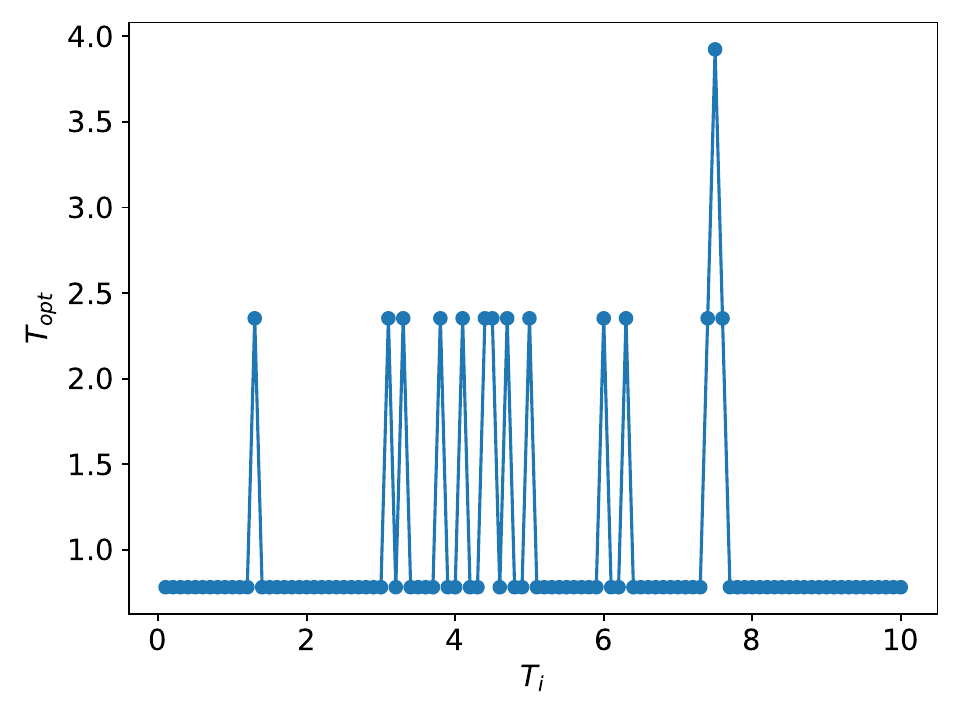}}
    \subfloat[\label{fig:optimal_T_hist_dipole_0.5_M_8}]{\includegraphics[width=0.32\textwidth]{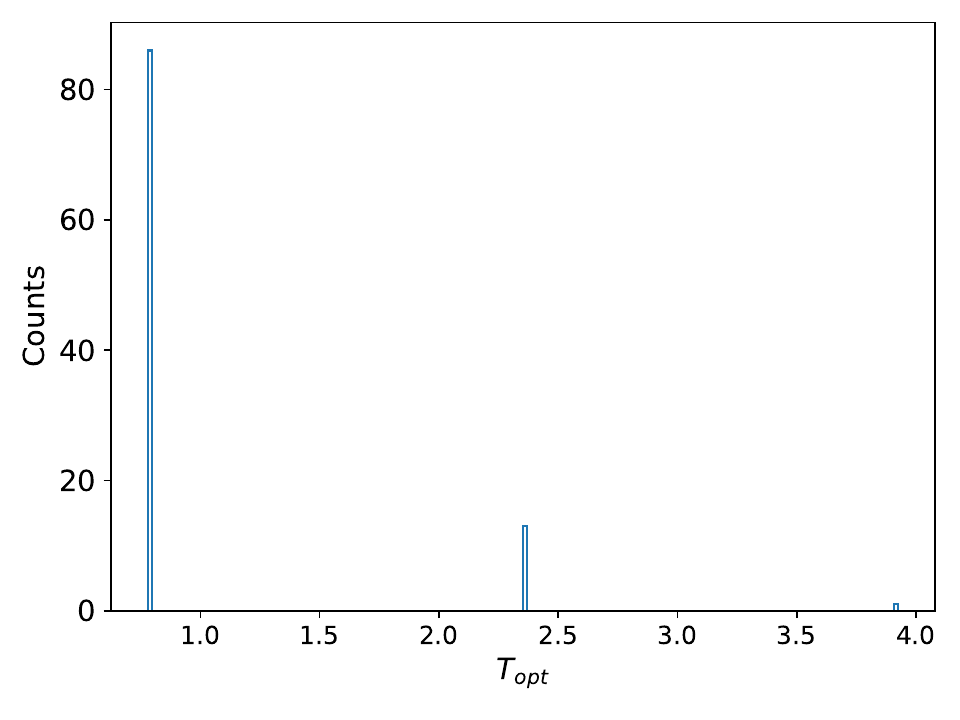}}
    \caption{Results of the gate compilation of CZ for $\Omega \gg J$ are shown: (a) The number of function evaluations, $nfev$, to reach the convergence, and (b) The resulting optimal time, $T_{opt}$, for each initial evolution time (CRAB)/ initial guess of optimal time (TCRAB), $T_{i}$. (c) The distribution of optimal time, $T_{opt}$, found by the TCRAB scheme.}
\label{fig:dipole_0.5_M_8}
\end{figure*}

In \cref{fig:bell_state_M_8,fig:lmg_M_8,fig:swap_1.0_M_8,fig:dipole_0.5_M_8}, we present the additional numerics we perform alongside the results in \cref{sec: results}.

\section{Oscillation of Optimised Fidelity}\label{sec:oscillation}
The Liouville superoperator of the unitary part can be explicitly split into a term $\mathcal{L}_{0}$ that corresponds to the drift Hamiltonian $H_0$ in \cref{eqn:Hamiltonian_TCRAB} and thus is independent of the evolution time $t$ and the control parameters $\vec{\alpha}$, and another term $\mathcal{L}_{C}$ that corresponds to the rest of the controlled Hamiltonian:
\begin{align*}
    \mathcal{L}_{H}(\vec{\alpha}, t) = \mathcal{L}_{0} + \mathcal{L}_{C}(\vec{\alpha}, t).
\end{align*}
When the drift Hamiltonian $H_0$ in \cref{eqn:Hamiltonian_TCRAB} commutes with all of the controlled Hamiltonian, then the evolution due to $\mathcal{L}_{0}$ and $\mathcal{L}_{C}(\vec{\alpha}, t)$ becomes separable and the final state can reach under the given Hamiltonian in the absence of noise can thus be written as:
\begin{align}
    \pket{\rho_{f}  (T, \vec{\alpha})} = e^{\mathcal{L}_{0}T} \pket{\rho_{c}  (T, \vec{\alpha})}
\end{align}
where $\rho_{c}$ is the state obtained under the evolution caused by  $\mathcal{L}_{C}(\vec{\alpha}, t)$, which is purely due to the control Hamiltonian, and $e^{\mathcal{L}_{0}T}$ is the action due to purely the drift Hamiltonian. 

In this way, the optimal fidelity for state-to-state transfer with evolution time $T$ can be written as:
\begin{align*}
    F_{\mathrm{opt}}(T) = F (T, \vec{\alpha}_T) & = \pbraket{\rho_{g,\mathrm{noi}}(T)}{\rho_{f}  (T, \vec{\alpha}_T)} \\
    & = \pbra{\rho_{g,\mathrm{noi}}(T)}e^{\mathcal{L}_{0}T} \pket{\rho_{c}  (T, \vec{\alpha}_T)}\\
    &= \Tr(\rho_{g,\mathrm{noi}}(T)e^{-i H_0 T} \rho_{c}(T, \vec{\alpha}_T) e^{i H_0 T})
\end{align*}

From here on, let us suppose $H_0$ is proportional to an involution operator $H_0 = \omega K_0$, i.e. it squares to $I$, which includes the Pauli operators. We then have $e^{\pm i H_0 T} = I \cos(\omega T) \pm iK_0\sin(\omega T)$ and
\begin{align*}
    F_{\mathrm{opt}}(T) & = \Tr(\rho_{g,\mathrm{noi}}\rho_{c}) \cos^2(\omega T) + \Tr(\rho_{g,\mathrm{noi}}K_0\rho_{c} K_0) \sin^2(\omega T) \\
    & - i [\Tr(\rho_{g,\mathrm{noi}}K_0\rho_{c}) - \Tr(\rho_{g,\mathrm{noi}}\rho_{c}K_0)] \sin(\omega T)\cos(\omega T)\\
    & = a(T) +  b(T) \cos(2 \omega T) + c(T) \sin(2 \omega T)
\end{align*}
with
\begin{align*}
    a(T) &= \frac{1}{2} \left(\Tr(\rho_{g,\mathrm{noi}}\rho_{c}) + \Tr(\rho_{g,\mathrm{noi}}K_0\rho_{c} K_0) \right)\\
    b(T)&= \frac{1}{2} \left(\Tr(\rho_{g,\mathrm{noi}}\rho_{c}) - \Tr(\rho_{g,\mathrm{noi}}K_0\rho_{c} K_0) \right)\\
    c(T) &= - \frac{i}{2} [\Tr(\rho_{g,\mathrm{noi}}K_0\rho_{c}) - \Tr(\rho_{g,\mathrm{noi}}\rho_{c}K_0)]
\end{align*}
For general $H_0$, we can still have oscillation, but more Fourier components will be involved~\cite{koczorQuantumNaturalGradient2022}.

However, commutation does not always mean there will be oscillation. For example, the control Hamiltonian can contain all of the basis of the drift Hamiltonian, which can then compensate for the effect of the drift Hamiltonian.

\section{Identity Test}\label{appendix:identity_test}

As described in \cref{sec: sensitivity}, one can perform the identity test: The test to check whether the evolution operator can be the same as the identity operator with the given drift and control Hamiltonian, i.e. $U(T, \vec{\alpha}) = \mathcal{I}$. The identity test checks the maximum capability of the control Hamiltonian to compensate for the drift term within the evolution operator, which causes oscillation in the infidelity. For example, if the infidelity of the identity test is $0.4$, this is the maximum capability of the control Hamiltonian as the control Hamiltonian cannot further suppress the drift term and make the infidelity lower. 

For a state-to-state transfer problem, the identity test reduces to the compilation of identity for the given initial state, i.e.$\abs{\bra{\psi_i}U(T, \vec{\alpha})\ket{\psi_i}}^2 = 1$. This shows that the time evolution operator, $U(T, \vec{\alpha})$, successfully acts like an identity operator for $\ket{\psi_i}$, but not necessarily for all states. In other words, for the given initial state, $\ket{\psi_i}$, the control Hamiltonian can fully suppress the oscillation by the drift term.

Note that this is a rough test of such capability, but this doesn't guarantee better performance in a specific problem, e.g. compilation of the CZ gate. This is because, depending on the target, one may need some effect of the drift term in addition to a specific form of the control pulse, which wouldn't be possible to obtain with the given set of basis functions and evolution time.

\begin{figure*}[htbp]
    \centering
    \subfloat[\label{fig:identity_test_bell_pair}]{\includegraphics[width=0.495\linewidth ]{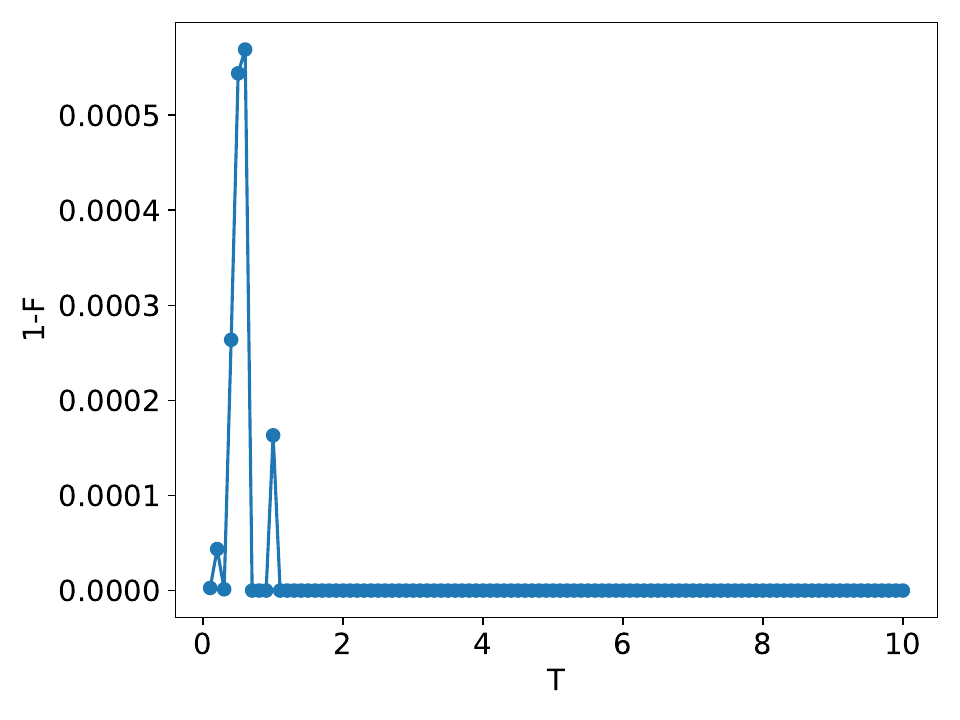}}~
    \subfloat[\label{fig:identity_test_LMG}]{\includegraphics[width=0.495\textwidth]{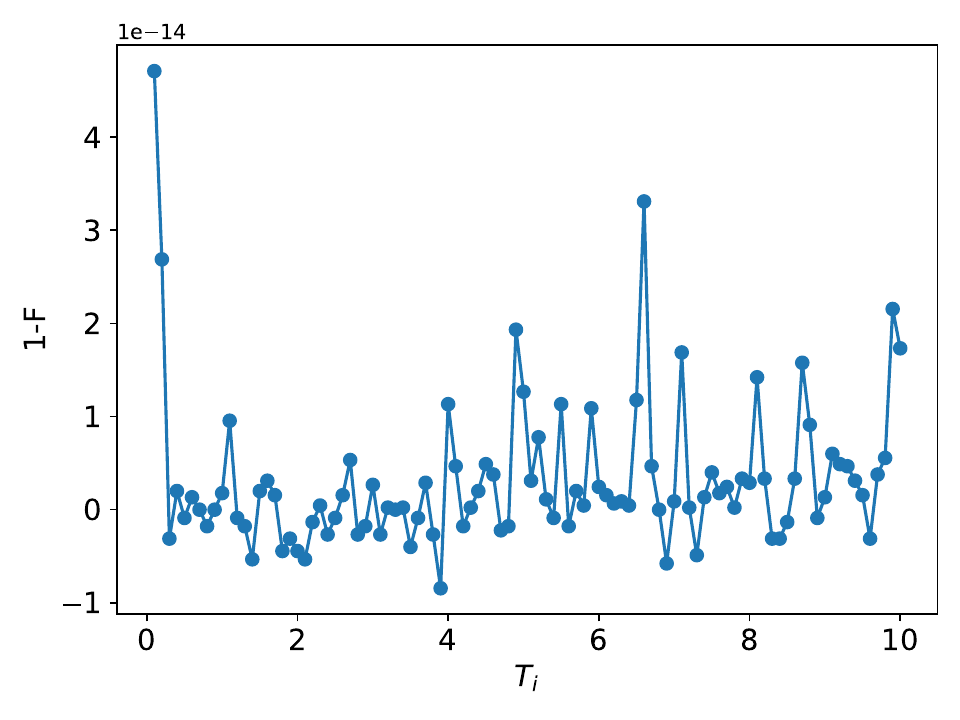}}
    \hfill
    \subfloat[\label{fig:identity_test_CZCompDipole}]{\includegraphics[width=0.495\textwidth]{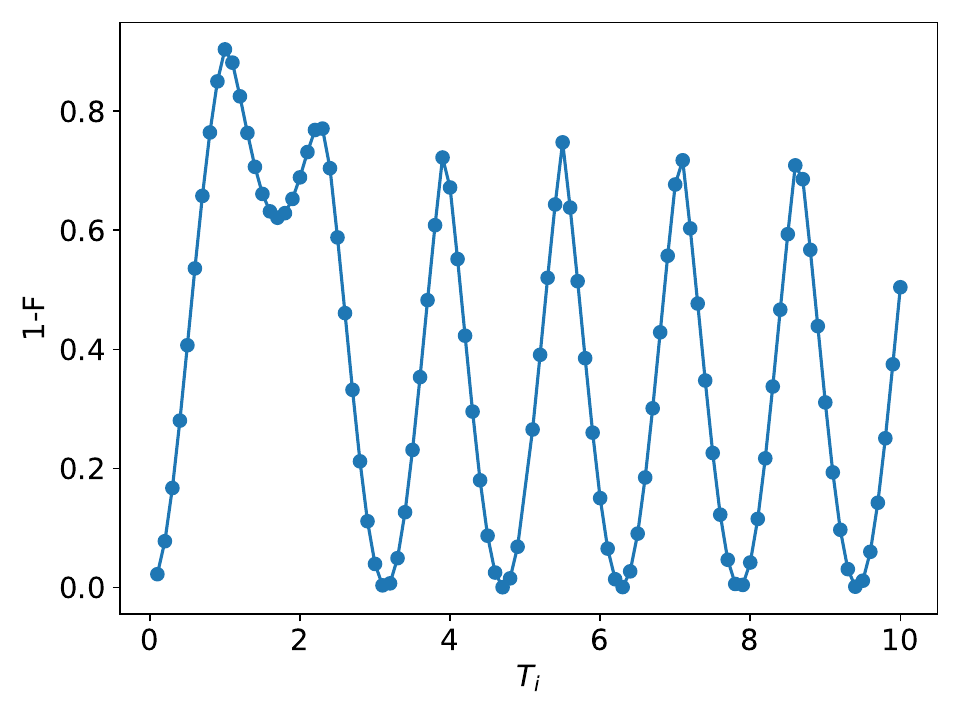}}~
    \subfloat[\label{fig:identity_test_CZCompGlobalZ}]{\includegraphics[width=0.495\textwidth]{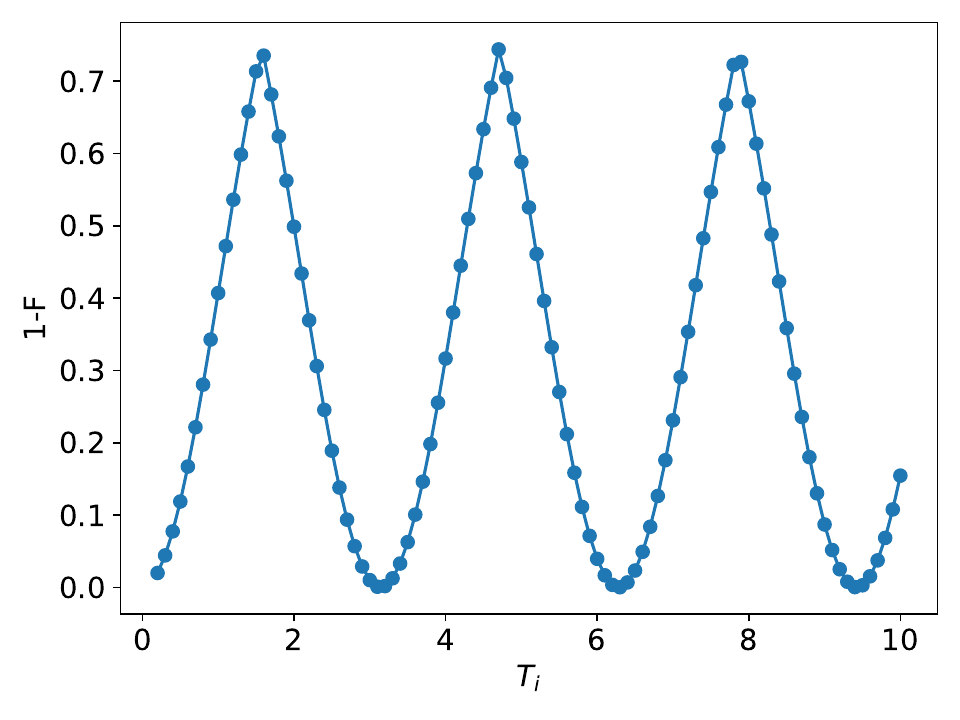}}

    \caption{Results of the identity test for (a) two capacitively coupled Josephson charge qubits, (b) the LMG model, and two spin qubits in Silicon quantum dots in the regime of (c) $\Omega \ll J$ and (d) $\Omega \gg J$. For all four cases, we performed CRAB with varying $T_{i}$ for the gate compilation of the identity gate using the given drift and control Hamiltonians. For a state-to-state transfer problem, we only need to show the compilation of identity for the given initial state. Thus, we used the initial states specified in \cref{subsec: results:state-to-state-transfer} and set the target states the same as the initial states. On the other hand, we used the Choi state scheme as explained for the gate compilation \cref{subsubsec:choi_state}.}
\label{fig:identity_test}
\end{figure*}

\cref{fig:identity_test} shows the results of the identity test for the four systems in \cref{sec: results}: (a) Two capacitively coupled charge qubits, (b) the LMG model, and two spin qubits in Silicon quantum dots of two regimes: (c) $\Omega \ll J$ and (d) $\Omega \gg J$.

The capability to suppress the effect of the drift term is determined by the commutation relation of the drift term and the control Hamiltonian. For a state-to-state transfer problem, this capability also depends on the initial state and the target state. 

The identity test for Josephson charge qubits in \cref{fig:identity_test_bell_pair} exhibits some peaks for low evolution times. While the control Hamiltonian, $\sigma^{z}_{1}\sigma^{z}_{2}$, anti-commutes with some parts of the drift term, i.e. $\sigma^{x}_{1}$ and $\sigma^{x}_{2}$, it commutes with the other half of the drift terms, i.e $\sigma^{z}_{1}$ and $\sigma^{z}_{2}$. The effect of commuting drift terms can still be compensated with the help of the anti-commuting drift Hamiltonian at the expense of longer evolution times. Nevertheless, The magnitudes of peaks are so small compared to the effects of decay and control terms, such that the cost function in \cref{fig:cost_func_bell_state} doesn't exhibit oscillation for short evolution time for the given initial and target states, i.e. $\ket{00}$.

For the LMG model in \cref{fig:identity_test_LMG}, the infidelity is in the order of $10^{-14}$ for all evolution times, suggesting that the control pulse can successfully suppress the contributions from the drift term if necessary for the given initial state, $\ket{000}$. Thus, the cost function in \cref{fig:cost_func_lmg} doesn't exhibit the oscillation.

Finally, the control Hamiltonians for spin qubits in Silicon quantum dots, i.e. SWAP for $\Omega \ll J$ and $Z_{1} \otimes Z_{2}$, commute with the drift term, $(\Delta E_{1}Z_{1} + \Delta E_{2} Z_{2})/2$, which is a sum of two single-qubit Z gates. Thus, there is no way for the control Hamiltonians to compensate for the oscillation due to the drift Hamiltonian, and the cost functions in \cref{fig:cost_func_dipole_0.5_M_8} and \cref{fig:cost_func_swap_1.0_M_8} exhibit oscillations.

\bibliography{refs}
\end{document}